\documentclass{elsart}

\usepackage{amssymb,latexsym, amsmath}
\usepackage{graphicx}
\usepackage{psfig}

\begin{document}

\begin{frontmatter}

\title{Lattice gas cellular automata model for rippling and
  aggregation in myxobacteria}

\author[ND]{Mark S. Alber}\ead{malber@nd.edu},
\author[LANL]{Yi Jiang}\ead{jiang@lanl.gov} \and
\author[ND]{Maria A. Kiskowski}\ead{mkiskows@nd.edu}

\address[ND]{Department of Mathematics and the
Interdisciplinary Center for the Study of Biocomplexity,
University of Notre Dame, Notre Dame, IN 46556-5670}
\address[LANL]{Theoretical Division, Los Alamos National Laboratory,
  Los Alamos, NM 87545}

\begin{abstract}

A lattice-gas cellular automaton (LGCA) model is used to simulate
rippling and aggregation in myxobacteria. An efficient way of
representing cells of different cell size, shape and orientation is
presented that may be easily extended to model later stages of
fruiting body formation. This LGCA model is designed to investigate
whether a refractory period, a minimum response time, a maximum
oscillation period and non-linear dependence of reversals of cells on
C-factor are necessary assumptions for rippling. It is shown that a
refractory period of 2-3 minutes, a minimum response time of up to 1
minute and no maximum oscillation period best reproduce rippling in
the experiments of {\it Myxoccoccus xanthus}. Non-linear dependence of
reversals on C-factor is critical at high cell density. Quantitative
simulations demonstrate that the increase in wavelength of ripples
when a culture is diluted with non-signaling cells can be explained
entirely by the decreased density of C-signaling cells. This result
further supports the hypothesis that levels of C-signaling
quantitatively depend on and modulate cell density. Analysis of the
interpenetrating high density waves shows the presence of a phase
shift analogous to the phase shift of interpenetrating
solitons. Finally, a model for swarming, aggregation and early
fruiting body formation is presented.

\end{abstract}


\end{frontmatter}

\section{Introduction}

Myxobacteria are one of the prime model systems for studying cell-cell
interaction and cell organization preceding differentiation.
Myxobacteria are social bacteria which swarm, feed and develop
cooperatively \cite{Kaiser1993}.  When starved, myxobacteria
self-organize into a three-dimensional fruiting body structure.
Fruiting body formation is a complex multi-step process of alignment,
rippling, streaming and aggregation that culminates in the
differentiation of highly elongated, motile cells into round,
non-motile spores.  A successful model exists for the fruiting body
formation of the eukaryotic slime mold {\it Dictyostelium discoideum}
\cite{Jiang,152,maree2001}. Understanding the formation of fruiting
bodies in myxobacteria, however, would provide a new insight since
collective myxobacteria motion depends not on chemotaxis as in {\it
Dictyostelium} but on contact-mediated signaling (see
\cite{dworkin1996} for a review).

\begin{figure}[ht]
\begin{center}
\includegraphics[height=2.0in]{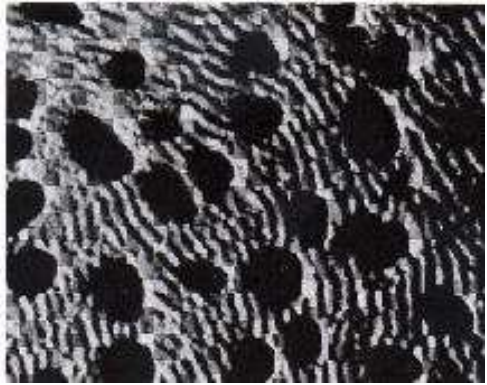}
\end{center}
\caption{\label{ripples} A field of immature fruiting bodies, shown as
  dark patches,  with ripples formed by cells outside of the
 aggregates. (From Shimkets and  Kaiser \cite{shimkets1982} with
 permission.) }
\end{figure}

During fruiting body formation myxobacteria cells are elongated, with
a 10:1 length to width ratio, and move along surfaces by
gliding. Gliding occurs in the direction of a cell's long axis
\cite{buchard1981} and is controlled by two distinct motility systems
in myxobacteria \cite{wolgemuth,hodgkin1979}. One of the most
interesting patterns that develops during myxobacteria morphogenesis
is rippling, which often occurs spontaneously and transiently during
the aggregation phase \cite{reichenbach1965,shimkets1982,welch2001}.
Rippling myxobacteria form equidistant ridges of high cell density
which appear to advance through the population as rhythmically
traveling waves \cite{reichenbach1965,shimkets1982} (Figure
\ref{ripples}). Cell movement in a ripple is approximately
one-dimensional since the majority of cells are aligned and move in
parallel lines with or against the direction of wave propagation
\cite{sager1994}.  Tracking individual bacteria within a ripple has
shown that cells reverse their traveling directions back and forth and
that each travels on the order of one wavelength between reversals
\cite{sager1994}. The ripple waves propagate with no net transport of
cells \cite{sager1994} and wave overlap causes neither constructive
nor destructive interference \cite{sager1994}. Although mechanisms for
gliding are not fully understood, they are believed to account for
both alignment \cite{wolgemuth,buchard1984} and reversals
\cite{wolgemuth} in myxobacteria.

Rippling is related to a membrane-associated signaling protein called
{\it C-factor}. C-factor regulates rippling
\cite{shimkets1982,sager1994,gronewold2001}, cells without the ability
to produce C-factor fail to ripple \cite{shimkets1982} and the
addition of C-factor (extracted from fruiting body cultures) causes
cell reversal frequencies to increase three-fold \cite{sager1994}.
C-signaling occurs via the direct cell-cell transfer of C-factor when
two elongated cells collide head to head
\cite{sager1994,kim1990_1,kim1990_2,kim1990_3,kroos1988}.
Understanding the mechanisms of the rippling phase may
reveal many clues about the way myxobacteria organize collective
motion since C-factor is also involved in all other stages of
fruiting body formation. For example, cells lacking in C-factor fail
to aggregate or sporulate \cite{hagen1978,shimkets1983,shimkets1990}
while high concentrations of exogeneous C-factor induce aggregation
and sporulation \cite{kim1990_2,shimkets1983,kim1991,li1992}.

In this paper we use two lattice gas cellular automata (LGCA) models
to simulate rippling and aggregation during the fruiting body
formation of myxobacteria, to show the potential of cellular automata
as models for biological pattern formation processes, and to evaluate,
in particular, the necessity of different biological assumptions shown in
previous models for pattern formation in myxobacteria. 

\begin{figure}[ht]
\begin{center}
\includegraphics[height=1.75in]{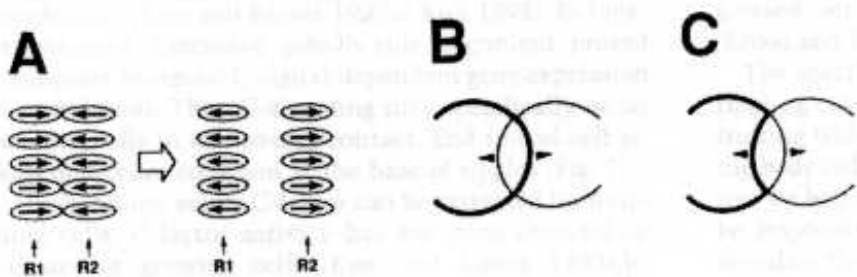}
\end{center}
\caption{\label{pr}
{(A) A reflection model for the interaction between individual cells
    in two counter-migrating ripple waves. Laterally aligned cells in
    counter-migrating ripples (labeled R1 and R2) reverse upon end to
    end contact. Arrows  represent the directions of cell
    movement. Relative cell positions are preserved.  (B) Morphology
    of ripple waves after collision. Thick and thin lines represent
    rightward and leftward moving wave fronts, respectively. Arrows
    show direction of wave movement. (C) Reflection of the same waves
    shown in B, with the ripple cell lineages modified to illustrate
    the effect of reversal. (From Sager and Kaiser \cite{sager1994}
    with permission).}}
\end{figure}

Sager and Kaiser \cite{sager1994} have proposed that precise
reflection explains the lack of interference between
wave-fronts for myxobacteria rippling.  Oriented collisions between
cells initiate C-signaling that causes cell reversals. According to
this hypothesis of precise reflection, when two wave-fronts collide,
the cells reflect one another, pair by pair, in a precise way
that preserves the wave structure in mirror image. Figure \ref{pr}
shows a schematic diagram of this reflection.

We present a new LGCA approach for modeling cells which is
computationally efficient yet approximates continuum dynamics more
closely than assuming point-like cells. As an example of this new
approach, we present a model for myxobacteria rippling based on the
hypothesis of precise reflection and a model for aggregation based on
C-signaling. 

This paper is organized as follows. The biological assumptions for
precise reflection and C-signaling that motivate the models are
described in the next section. In section 3 we describe specifics of
two LGCA models. In section 4 results of modeling rippling phenomenon
are discussed in detail. Section 5 provides description of a model for
aggregation centers. The paper ends with a summary section.

\section{Biological Background}
In this section we describe the biological observations which
motivate our models for rippling and aggregation.

Rippling and aggregation are both controlled by C-signaling
and are characterized by specific high cell density patterns (in
particular, moving high density ridges in rippling and stationary high
density mounds in aggregation). There is a marked relationship between
cell density, levels of C-signaling and behaviors in myxobacteria
triggered by C-signaling \cite{kim1992}. 
C-signaling increases with density since end-to-end contacts
between cells are more likely with increased density
\cite{sager1993,julien2000} and high cell densities favor spatial
arrangements in which there are many end-to-end contacts due to
the polarity of myxobacterial cells \cite{sager1993,julien2000}.
Cell density and C-signaling levels increase together from
rippling to aggregation and from aggregation to sporulation
\cite{sager1993,julien2000}. Further, increased thresholds of
C-factor induce rippling, aggregation and sporulation respectively
\cite{kim1991,li1992,kruse2001}, suggesting C-signaling levels, as
a measure of cell density, are checkpoints for different stages of
development. Kim and Kaiser suggest that C-factor may act as a
developmental timer that triggers sporulation only when cell density
is as high as possible \cite {kim1990_3,kim1991}. A high density
aggregate will culminate in a fruiting body with a large number of
spores ensuring that the next cycle is started by a population of
cells \cite{reichenbach1993}.  Sager and Kaiser have also observed
the effect of C-signaling-competent cell density upon ripple
wavelength \cite {sager1994}. They dilute a cell population of
C-signaling-competent cells with cells that are able to respond to
C-factor but are not able to transmit it. They find that with
increased concentrations of these csgA-minus cells, ripple
wavelength increases non-linearly.

In addition to cell density patterns, cell geometries are
important throughout the stages of fruiting body formation and
distinguish different stages. During the fruiting body formation,
cells form aligned patches from a random distribution
\cite{kaiser1999}.  For rippling, a large number of cells must be
aligned both parallel and anti-parallel within the same field. For
streaming, cells form long chains which flow cooperatively in
aggregation centers \cite{jelsbak2000}. In {\it Stigmatella spp.},
cells moving in circles or spirals form microscopic transient
aggregates.  These aggregates disappear as cells also spiral away
tangentially \cite{white1993}. Macroscopic aggregates form in
areas of high density \cite{white1993} and may also disappear as
cells apparently stream along chains from one aggregate to the
other \cite{kaiser_private}. The mature structures of fruiting
bodies are diverse and species-dependent, ranging in size between
10 and 1000 $\mu$m \cite{reichenbach1993}. In {\it Myxococcus
xanthus}, the basal region of the fruiting body is a shell of
densely packed cells which orbit in two directions, both
clock-wise and counter-clockwise, around an inner region only
one-third as dense \cite{sager1993,julien2000}. In {\it
Stigmatella} aggregates, cells are organized in concentric circles
or ellipses and cells move in a spiral fashion up the aggregate as
the fruiting body develops \cite{white1993,vasquez1985}.

Current models for rippling (\cite{br,igoshin,lutscher2002b})
assume precise reflection.  Key differences among these models include
their biological assumptions regarding the existence of internal
biochemical cell cycles.  It is still not known if an internal cell
timer is involved in myxobacterial rippling.  Several models with 
completely different assumptions all qualitatively produce ripple
patterns resembling experiment.

An {\it internal timer} is a hypothetical molecular cell clock
which regulates the interval between reversals. The internal timer
may specify a delay, or minimum period between reversals, which
would include the refractory period, see below, {\it and} a
minimum response time; the minimum period of time required for a
non-refractory cell to become stimulated to turn. Also, the
internal timer may specify a maximum oscillation period, in which
case the timer may speed up or slow down depending upon
collisions, but the cell will always  turn within a specified
period of time even without collisions. Individual pre-rippling
cells reverse spontaneously every 5-10 minutes with a variance in
the period much smaller than the mean
\cite{igoshin,jelsbak1999,shi1996}. This would suggest that there 
is a component of the timer specifying a maximum oscillation
period.  Also, observation of rippling bacteria reveals that cells
oscillate even in ripple troughs where the density is too low for
frequent collisions \cite{sager1994} further supporting the
hypothesis of a maximum oscillation period.

The {\it refractory period} is a period of time immediately
following a cell reversal, during which the cell is insensitive to
C-factor. Although there is no evidence of  a refractory period in
the C-signaling system, the refractory period is a
general feature of bacterial signaling systems \cite{igoshin} (for
a description of the role the refractory period plays in {\it
Dictyostelium}, see \cite{goldbeter}). The addition of .02 units of
external C-factor triples the reversal frequency of single cells
from .09 reversals per minute to .32 reversals per minute
\cite{sager1994}. Cells do not reverse more frequently at still
higher levels of C-factor, however, suggesting the existence of a
minimum oscillation period of 3 minutes in response to C-factor.
This minimum oscillation period would be the sum of the refractory
period and the minimum response, so the duration of the
refractory period cannot be guessed from this fact alone.

To resolve the conflicts of these models for rippling our first
LGCA model is designed to test different assumptions. The results of
our model for rippling shows that rippling is stable for a wide
range of parameters, C-signaling plays an important role in
modulating cell density during rippling, and non-C-signaling cells
have no effect on the rippling pattern when mixed with wild-type
cells. Further, by comparing model results with experiments, we can
conclude reversals during rippling would not be 
regulated by a built-in maximum oscillation period.

We then present a second LGCA model for aggregation based on
C-signal alignment, which reproduces the sequence and geometry of
the non-rippling stages of fruiting body formation in detail,
showing that a simple local rule based on C-signaling can account
for many experimental observations.

\section{Model and Method}

LGCA are relatively simple Cellular Automata models.  They employ a
regular, finite lattice and include a finite set of particle states,
an interaction neighborhood and local rules that determine the
particles' movements and transitions between states \cite {dormann}.
LGCA differ from traditional CA by assuming particle motion and an
exclusion principle. The connectivity of the lattice fixes the number
of allowed non-zero velocities {\it or channels} for each
particle. For example, a nearest-neighbor square lattice has four
non-zero allowed channels. The channel specifies the direction and
magnitude of movement, which may include zero velocity (resting). In a
simple exclusion rule, only one particle may have each allowed
non-zero velocity at each lattice site. Thus, a set of Boolean
variables describes the occupation of each allowed particle state:
occupied (1) or empty (0). Each lattice site on a square lattice can
then contain from zero to four particles with non-zero velocity. 

The transition rule of an LGCA has two steps. An interaction step
updates the state of each particle at each lattice site. Particles may
change velocity state, appear or disappear in any number of ways as
long as they do not violate the exclusion principle. In the transport
step, cells move synchronously in the direction and by the distance
specified by their velocity state. Synchronous transport prevents
particle collisions which would violate the exclusion principle (other
models define a collision resolution algorithm). LGCA models are
specially constructed to allow parallel synchronous movement and
updating of a large number of particles \cite{dormann}.

\subsection{Representation of cells}

In classical LGCA, biological cells are dimensionless and represented
as a single occupied node on a lattice (e.g., see \cite{br} and
\cite{lutscher2002b}). Interaction neighborhoods are typically
nearest-neighbor or next-nearest-neighbor on a square lattice. The
exclusion principle makes transport unwieldy when a single
 cell occupies more than one node since a cell may only advance if
all the channels it would occupy are available.  Similarly, it is
difficult to model the overlapping and stacking of cells. Cells
without dimension are untenable for a sophisticated model of
myxobacteria fruiting body formation, however. Cells are very
elongated during rippling, streaming and aggregation and form regular,
dense arrays by cell alignment. Also, a realistic model of cell
overlap and cell stacking is needed since interaction occurs only at
specific regions of highly elongated cells and cell density is a
critical parameter throughout this morphogenesis.

B\"{o}rner {\it et al.} \cite {br} have mediated the problem of
stacking by introducing a semi-three-dimensional lattice where a third
z-coordinate gives the vertical position of each cell when it is
stacked upon other cells. Stevens \cite{stevens} has introduced a
model of rod-shaped cells that occupy many nodes and have variable
shape in her cellular automata model of streaming and aggregation in
myxobacteria. Neither of these two models are LGCA since they do not
incorporate synchronous transport along channels. We device a novel
way of representing cells which facilitates variable cell shape, cell
stacking and incomplete cell overlap while preserving the advantages
of LGCA; namely, synchronous transport and binary representation of
cells within channels (e.g., a `0' indicating an unoccupied channel
and a `1' indicating an occupied channel).

We represent the cells as (1) a single node which corresponds to the
position of the cell's center (or ``center of mass'') in the $xy$
plane, (2) the choice of occupied channel at the cell's position
designating the cell's orientation and (3) a local neighborhood
defining the physical size and shape of the cell with associated
interaction neighborhoods (Figure \ref{nbhds}).  The interaction
neighborhoods depend on the dynamics of the model and need not exactly
overlap the cell shape. In our models for rippling and aggregation, we
define the size and shape of the cell as a $3 \times \ell$ rectangle,
where $\ell$ is cell length.  As $\ell$ increases, the cell shape
becomes more elongated.  A cell length of $\ell=30$ corresponds to the
$1 \times 10$ proportions of rippling {\it Myxococcus xanthus} cells
\cite{kim1990_3}. Representing a cell as an oriented point with an
associated cell shape is computationally efficient, yet approximates
continuum dynamics more closely than assuming point-like cells, since
elongated cells may overlap in many ways. We have also solved the cell
stacking problem, since overlapping cell shapes correspond to cells
stacked on top of each other.  This cell representation conveniently
extends to changing cell dimensions and the more complex interactions
of fruiting body formation.

\subsection{LGCA model for rippling}

We assume precise reflection and investigates the roles of a cell
refractory period, a minimum response period, a maximum oscillation
period and non-linear dependence of reversals on C-factor
independently. 

\subsubsection{Local Rules}

\begin{enumerate}

\item{Our model employs a square lattice with periodic boundary
    conditions imposed at all four edges. Unit velocities are allowed 
    in the positive and negative $x$ directions. (A resting channel
    may be easily added to model a small percentage of resting cells
    as in \cite{br}.}

\item{Cells are initially randomly distributed with density $\delta$,
    where $\delta$ is the total cell area divided by total lattice area.}

\item{ Every cell is initially equipped with an internal timer by
    randomly assigning it a clock value between 1 and a maximum clock
    value $\tau$. We define a refractory period $R$ such that $0 \leq
    R < \tau$ (see a detailed description of the internal timer,
    below).  If the internal timer $\phi$ of a cell is less than $R$,
    the cell is refractory. Otherwise, the cell is sensitive.}

\item {At each time-step, the internal timer of each refractory cell
    is increased by 1 while the internal timer of sensitive cells is
    increased by an amount proportional to the number of head-on
    cell-cell collisions $n$ occurring at that timestep.} 

\item{When a cell's internal timer has increased past $\tau$, the cell
    reverses, the internal timer resets to 0, and the cell becomes
    refractory. Reversals occur as a cell's center switches
    from a right- or left-directed channel to a left- or
    right-directed channel, respectively.}

\item{During the final transport step, all cells move synchronously
    one node in the direction of their velocity by updating the
    positions of their centers. Separate velocity states at
    each node ensure that more than one cell never occupies a single
    channel.}

\end{enumerate}

\subsubsection{Internal timer}

We model an internal timer with three parameters; $R$, $t$ and $\tau$.
$R$ is the number of refractory time-steps, $t$ is the minimum number
of time-steps until a reversal and $\tau$ is the maximum number of
time-steps until a reversal. The minimum period of time required for a
sensitive cell to become stimulated to turn is the minimum response
period $t-R$. During the refractory period, cells are insensitive to
collisions and the internal timer advances at a uniform rate.
After the refractory period, cells become sensitive and during this phase
the number of head-on cell-cell collisions accelerates the internal timer
so that the interval between reversals shortens. This acceleration is
density-dependent, so that many simultaneous collisions accelerate the
internal timer more than only one collision.

Our internal timer extends the timer in Igoshin {\it et al.} \cite
{igoshin}. They used a phase variable $\phi$ to model an oscillating
cycle of movement in one direction followed by a reversal and movement
in the opposite direction. During the refractory period the phase
variable advances at a constant rate but during the sensitive period,
the phase variable advance may increase non-linearly with the number
of collisions.  Thus, the evolution of our timer determines reversal
rather than a collision as in the model of B\"{o}rner {\it et
al.} \cite{br}. The state of our internal timer is specified by $0\leq \phi(t)
\leq \tau$. $\phi$ progresses at a fixed rate of one unit per
time-step for $R$ refractory time-steps, and then progresses at a rate
$\omega$ that depends non-linearly on the number of collisions $n$ 
which have occurred at that timestep to
the power $p$:

\begin{equation}
\omega(x,\phi,n,q) = 1 +\left(\frac{\tau-t}{t-R}\right)*\left(
\frac{[min(n,q)]^p}{q^p}\right)*F(\phi),
\end{equation}

\noindent{where,}

\begin{equation}
F(\phi)=
\begin{cases}
  0,    &\text{for $0\leq\phi\leq R$;}\\ 0,    &\text{for
  $\pi\leq\phi\leq(\pi+R)$;}\\  1,    &\text{otherwise.}\\
\end{cases}
\end{equation}

This equation is the simplest that produces a reversal period of
$\tau$ when no collisions occur, a refractory period of $R$ time-steps
in which the phase velocity is one, and a minimum reversal period of
$t$ when a threshold (quorum) number $q$ of collisions occurs at every
sensitive time-step. There is ``quorum sensing'' in that the clock
velocity is maximal whenever the number of collisions at a time-step
exceeds the quorum value $q$. A particle will oscillate with the
minimum reversal period only if it reaches a threshold number of
collisions during each non-refractory time-step (for ($t-R$)
time-steps). If the collision rate is below the threshold, the clock
phase velocity is less than maximal. However, as the number of
collisions increases from 0 to $q$, the phase velocity increases
non-linearly as $q$ to the power $p$.

While in the model of B{\" o}rner {\it et al.} \cite{br} there is no minimum
response period for a cell to reverse, and in the model of Igoshin
{\it et al.} \cite{igoshin} a minimum response time is an inherent
component of the 
internal clock, our model incorporates ``on-off switches'' for a
refractory period, minimum response period, maximum oscillation period
and quorum sensing. Setting the refractory period equal to 0
time-steps in our model is the off-switch for the refractory period,
and setting $t=R+1$ is the off-switch for the minimum response
time. No maximum oscillation period is modeled by choosing a maximum
oscillation period $\tau$ greater than the running period of the
simulation, so that the automatic reversal of cells within $\tau$
time-steps has no effect on the dynamics of the simulation. There is
no quorum sensing if $q$ is set to 1 so that a single collision during
a timestep has the same effect as many collisions.

If there is no refractory period, cells are always sensitive to
collisions. If there is no minimum response time, cells may reverse
immediately after becoming sensitive if there are sufficiently many
collisions in one timestep. Finally, if there is no maximum
oscillation period, cells may never reverse without sufficiently many
collisions.

\subsubsection{Head-on cell-cell collisions}

\begin{figure}[ht]
\begin{center}
\includegraphics[height=.75in]{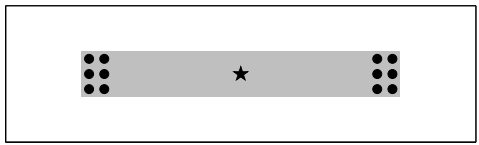}
\end{center} \caption{\label{nbhds}
The shaded rectangle corresponds to the cell shape of a right or left
moving cell in our model for rippling. This cell is $3\times 21$ nodes
for a $1\times 7$ aspect ratio. The star in this figure corresponds to
the cell's center and the nodes of the interaction
neighborhood where C-factor is exchanged are indicated by black
squares at the cell poles.}
\end{figure}

We define an interaction neighborhood of eight nodes for the exchange
of C-factor at the poles of a cell of length l (see Figure
\ref{nbhds}). The cell width of 3 nodes is larger than 1 to account
for coupling in the y-direction and the interaction neighborhood must
extend at least two nodes along the length of the cell to compensate
for the discretization of the lattice since cells traveling in
opposite directions may pass without their poles exactly overlapping.

A head-on cell-cell {\it collision} is defined to occur when the interaction
neighborhoods of two anti-parallel cells overlap. A cell may
collide with multiple cells simultaneously since the
interaction neighborhood is four nodes at each pole.  Note that the specific
shape of the cell is not important for rippling dynamics since the two
areas of C-signaling are the only places where interaction
occurs. Nevertheless, a shape extending over several nodes is
necessary to permit the necessary overlapping and stacking at high
density since the exclusion principle mandates that each channel has
at most one cell center.  Thus, the cell centers of two colliding cells
will be separated by one cell length and do not compete for channels at 
the same node. Also, for sufficiently long
cell lengths, the probability of more than one cell center located at
the same node is low even when the local cell density may be high.

We are able to simulate a rippling population with arbitrary
concentrations of both wild-type and non-C-signaling cells and
quantitatively reproduce their experimental results in detail, as did
Igoshin {\it et al.} using their continuum model
\cite{igoshin}. Further, we demonstrate that the change in wavelength
may be entirely explained by the change in density of C-signaling
cells.

\section{Rippling results and discussion}

\begin{figure}[ht]
  \begin{center}
    \includegraphics[height=1.5in]{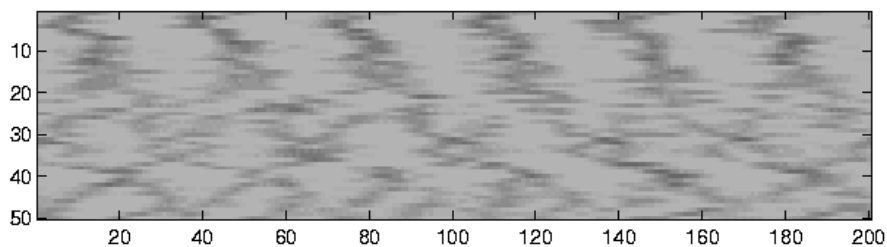}
  \end{center} 
  \caption{\label{ripples.eps} Typical ripple
    pattern including both a cell clock and refractory period in the
    model. (Cell length = 5, $\delta = 2$, $R=10$, $t=15$, $\tau=25$.)
    Figure shows the density of cells (darker gray indicates higher
    density) on a $50 \times 200$ lattice after 1000 time steps,
    corresponding to approximately 200 minutes in real time.}
\end{figure}

Our model forms a stable ripple pattern from a homogeneous initial
distribution for a wide range of parameters, with the ripples
apparently differing only in ripple wavelength, ripple density and
ripple width (see Figure \ref{ripples.eps}).

Absence of a maximum oscillation period is modeled by choosing a maximum
oscillation period $\tau$ greater than the running period of the
simulation, so that the automatic reversal of cells within $\tau$
time-steps has no effect on the dynamics of the simulation. We find
that ripples form with or without a maximum oscillation period over
the full range of densities. When there is a maximum oscillation
period, the maximum oscillation period must be chosen greater than
twice the refractory period for the development of ripples. There is
no upper bound on the maximum oscillation time, which is why the
maximum reversal period is unnecessary. Ripples develop most quickly
and cell oscillations are most regular {\it with} an internal timer
when the maximum oscillation period is carefully chosen with respect
to the other parameters of the model. Nevertheless, it appears that
experimental results are best reproduced when there is no maximum
oscillation period. 

A refractory period is required for rippling for cells of length
greater than 2 or 3 nodes, and although there may exist a minimum
response time of more than one time-step, it is an interesting result
of our model that the minimum response period $T-R$ must be small compared
to the refractory period. In particular, rippling occurs whenever the
minimum oscillation time $t$ is greater than $\ell/v$ time-steps  and the
refractory period $R$ is at least two-thirds $t$. The first condition
is required because if the minimum oscillation period $t$ is less than
the period of time it takes a cell to travel one cell length, two
cells or a cluster of cells will stimulate each other to oscillate in
place.  The second condition that the refractory period is at least
two-thirds the minimum oscillation period indicates that the minimum
response time of a cell can not be too long compared to the refractory
period.

Experiments suggest that the minimum oscillation period of a cell in
response to C-factor is about 3 minutes \cite{sager1994}. According to our result that
the minimum response time cannot be more than two-thirds the
refractory period, we can predict the existence of a refractory period
in myxobacteria cells, with a duration of 2-3 minutes.

\begin{figure}[ht]
\begin{center}
\includegraphics[height=1.8in]{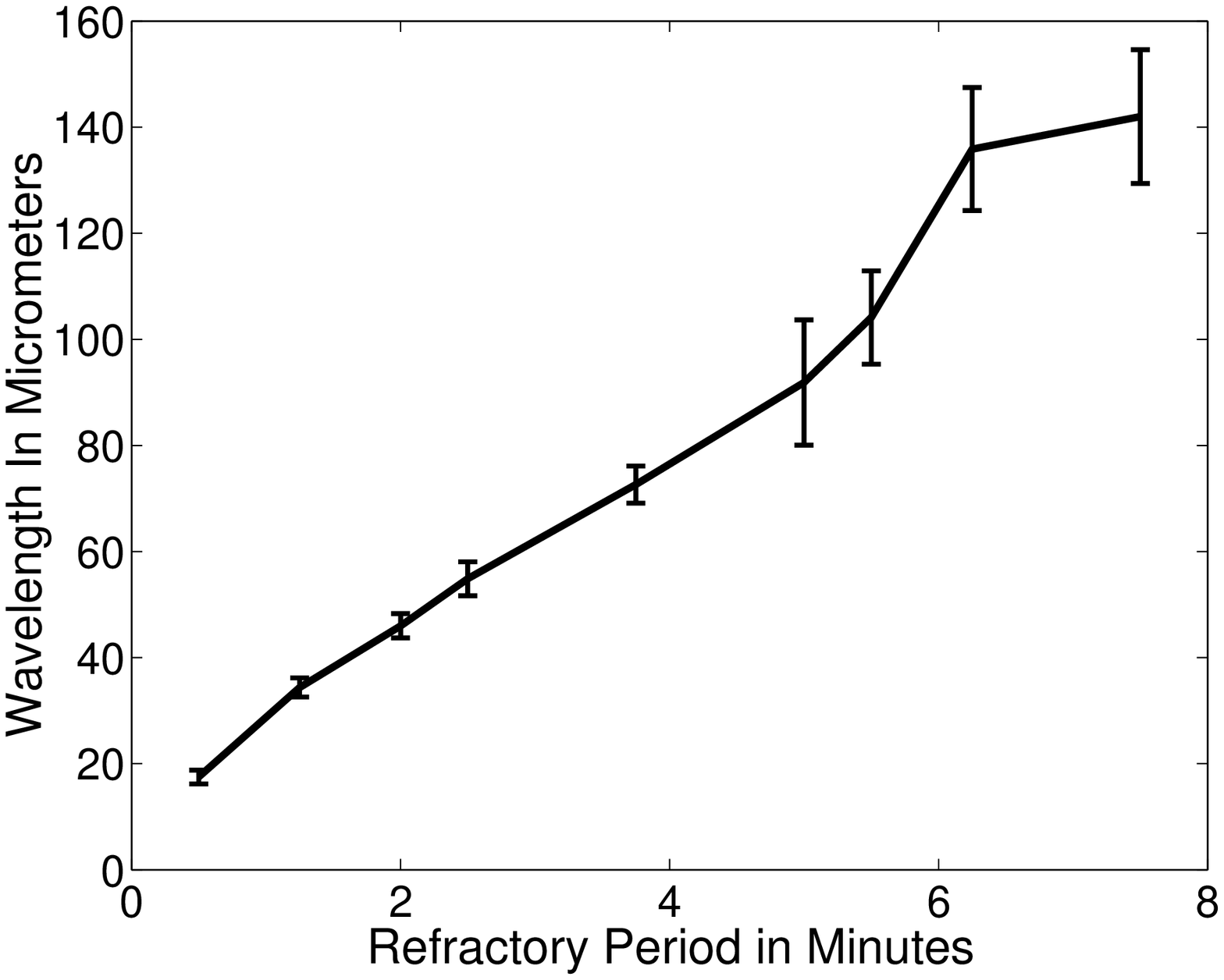}{ (a) }
\includegraphics[height=1.8in]{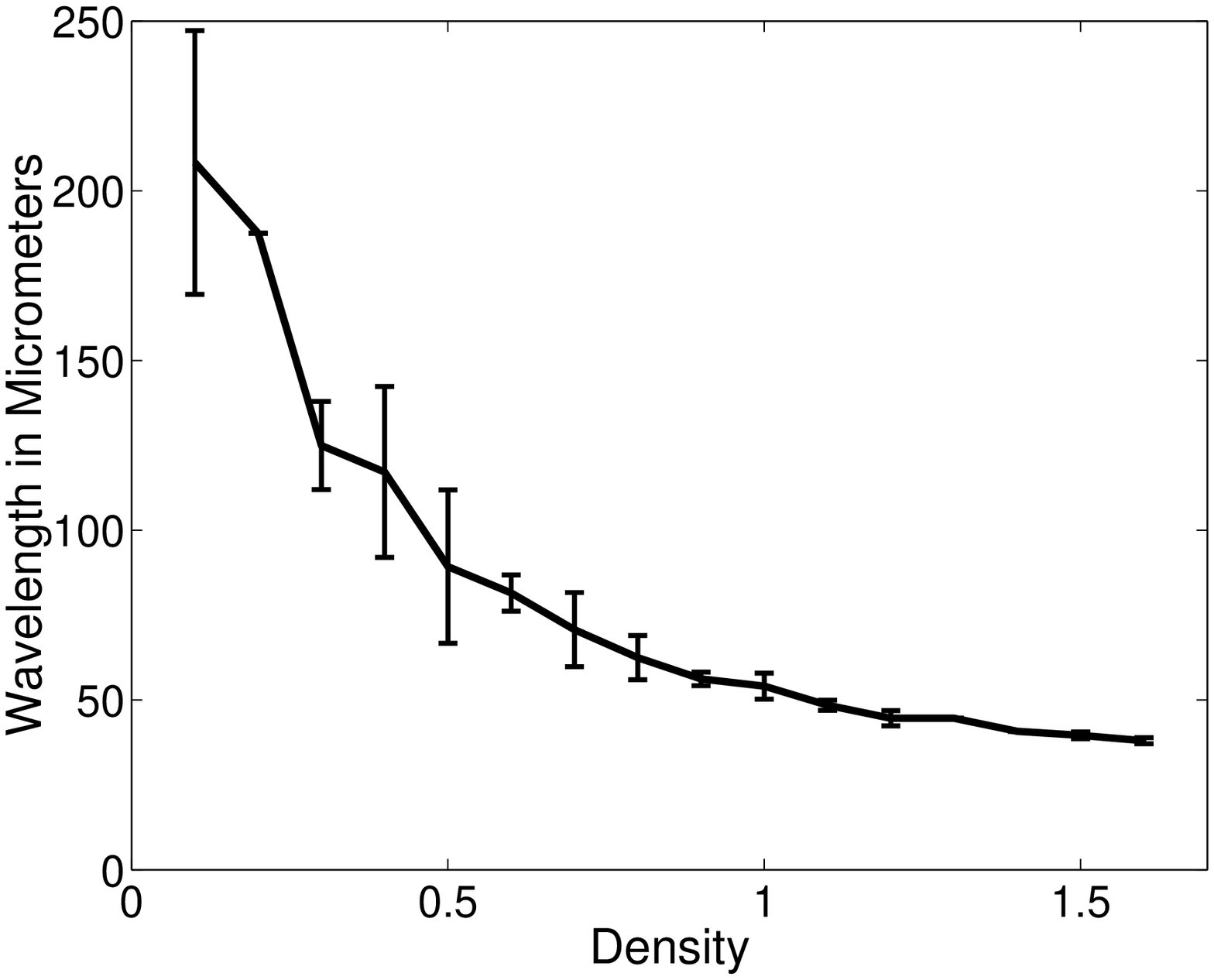}{ (b)}
\end{center}
\caption{\label{RDwavelength.eps}
a) Average wavelength in micrometers versus refractory period in
 minutes.  Cell length $\ell=4$, $\delta=1$. The internal timer is
 adjusted so that the fraction of clock time spent in the refractory
 period is constant: $t=3R/2$ and $\tau=5*R/2$. b) Average wavelength
 in micrometers versus density (total cell area over total lattice
 area). Cell length $=4$ with an internal timer given by $R=8$,
 $t=12$, $\tau=20$. }
\end{figure}

The wavelength of the ripples depends on both the duration of the
refractory period and the density of signaling cells. Figure
\ref{RDwavelength.eps} shows that the ripple wavelength increases with
increasing refractory period (a) and decreases with increasing cell
density (b). Notice that error bars that show standard deviations of
the mean wavelength over five simulations increases with wavelength.
A refractory period of 2-3 minutes yields a ripple wavelength of
about 60 micrometers (Figure \ref{RDwavelength.eps}a), which
corresponds well to typical experimental ripple wavelengths
\cite{sager1994}. The correspondence between refractory period and
wavelength given in Figure \ref{RDwavelength.eps} is a only rough
estimate, however.  We believe the reasons are that in these
simulations the cell density is relatively low, which decreases the
density of C-factor relative to experimental conditions, and cells are
not very elongated, which increases the density of C-factor relative
to experimental conditions.  

Note that in Figure \ref{RDwavelength.eps}a, the curve has a
wavelength of approximately 20 micrometers when the refractory period
is less than 1 minute. Since cells have a length of 5 micrometers,
this is the smallest wavelength that may be resolved as there is only
one cell length between subsequent high density waves. At very high
density, when the refractory period is 0, cells may be stimulated to
reverse every timestep, so that there would be, theoretically, a
wavelength of only 1 node. However, cells will be uniformly
distributed in this case and there will be no well-defined
high-density waves. In the simulations described in Figure
\ref{RDwavelength.eps}b, density is increased while refractory and
minimum oscillation periods have a constant value. The minimum
possible wavelength in this case is limited by the minimum oscillation
period. In particular, the minimum possible wavelength is twice the
minimum distance traveled by a cell between reversals, which is twice
the distance traveled during the minimum oscillation period, which is
30 micrometers in this example. Thus, even as density is increased
very high, the curve must have a horizontal asymptote at wavelength =
30 micrometers.

\subsection{Non-linear response of reversals to C-factor.}

Reversals depend on the number of collisions a cell encounters which
depends on the density of C-factor. Thus the number of collisions
required for a reversal, the quorum value $q$, should be a function of
the density of C-signaling nodes.  The density of C-signaling nodes is
a function of both cell density and cell length since longer cells
have a reduced C-signaling area to non-C-signaling area ratio.  Thus,
we describe optimal quorum values $q$ as a function of
C-signaling node density rather than cell density.

At a low density of C-signaling nodes, ripples form
even when both $q$ and $p$ are 1 so that only 1 collision during
the sensitive period is needed to trigger an reversal.
When the density of C-signaling nodes is
greater than or equal to 1, however, the chances of collisions are so high
in the initial homogeneous population that cells almost always reverse in the
minimum number of timesteps and, with no
differential behavior among cells, a rippling pattern fails to
form. Ripples will form at arbitrarily large densities of cells and
C-signaling nodes if the number of collisions needed to
trigger a reversal is increased. When the number of collisions
required for a reversal is greater than 3 ($q>3$), ripples develop
more quickly if the non-linear response to density $p$ is
increased greater than 1. A value of $p=3$ yields optimal rippling
for all quorum values and densities, which is consistent with the
results of Igoshin {\it et al.} \cite{igoshin} for their
value of $p$ \cite{igoshin}.

\subsection{Ripple phase shift}

\begin{figure}[ht]
    \begin{center}
        \includegraphics[height=1in]{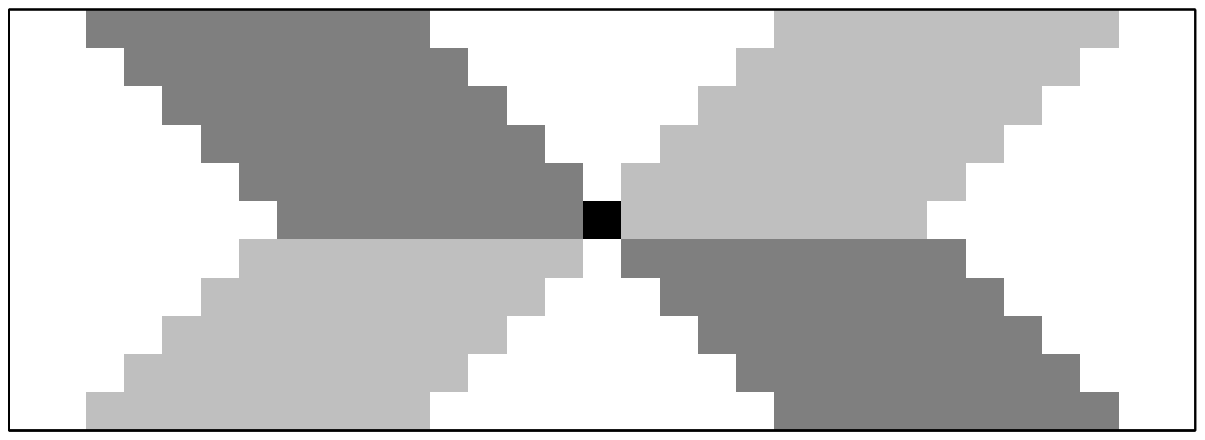}{ (a) }
        \includegraphics[height=1in]{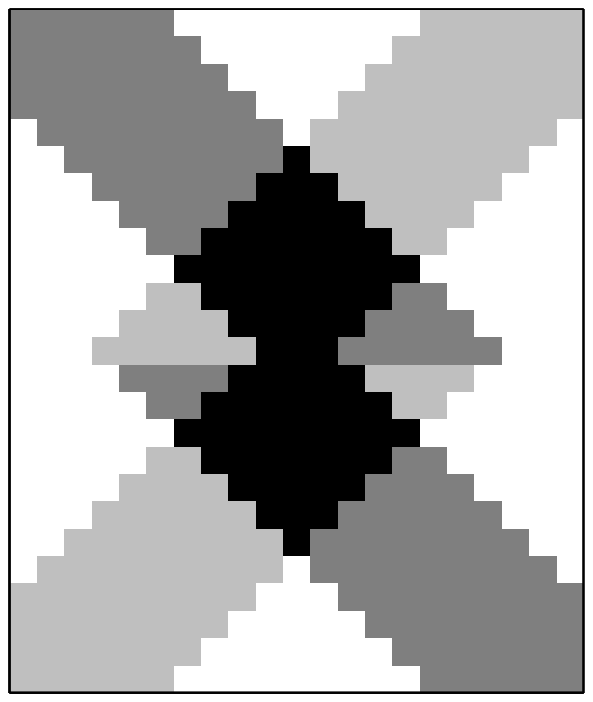}{ (b) }
    \end{center}
    \caption{\label{jumps} Space-time plot of a wave
        inter-penetration. Time increases as the vertical axis
        descends. Right-directed particles are shown in dark gray,
        left-directed particles are shown in light gray. a) Phase jump
        of one cell length (9 units) as two cells collide and
        immediately reverse. b) Phase delay as two cells collide and
        travel 8 time-steps before reversing.}
\end{figure}

Counter-propagating ripples appear to pass through each other with no
interference, which lead Sager and Kaiser to propose the hypothesis of
precise reflection \cite{sager1994}. Indeed, tracking of
right-propagating ripples and left-propagating ripples in Figure
\ref{solitons2}a, shows that the waves move continuously despite
collisions and subsequent reflection. Inspection of the collision and
subsequent reversal of two cells, however, shows there is a jump in
phase equal to exactly one cell length if they reverse immediately
upon colliding (see Figure \ref{jumps}a). This phase jump occurs
because a cell reverses by changing its orientation rather than by
turning: when a right-moving particle collides with a left-moving
particle and reverses, it is exactly one cell length ahead of the
left-moving cell that it replaces. When all of the particles within a
ripple are in phase, as is often the case, this jump is also seen in
the ripple waves as two waves interpenetrate. If the cells continue
$p$ more steps before reversing (for example, if their clocks were
almost near $\tau$ after the collision), then there would be a phase
jump of $\ell-2p$. If $2p>\ell$, there will be a phase delay (see Figure
\ref{jumps}b).  In their continuous model, Igoshin {\it et al.}
(\cite {igoshin}, Figure 3b) also showed when ripples collide a small
jump in phase reminiscent of a soliton jump. 

\subsection{Effect of dilution with non-signaling cells}

\begin{figure}[ht]
\begin{center}

\includegraphics[height=1.8in]{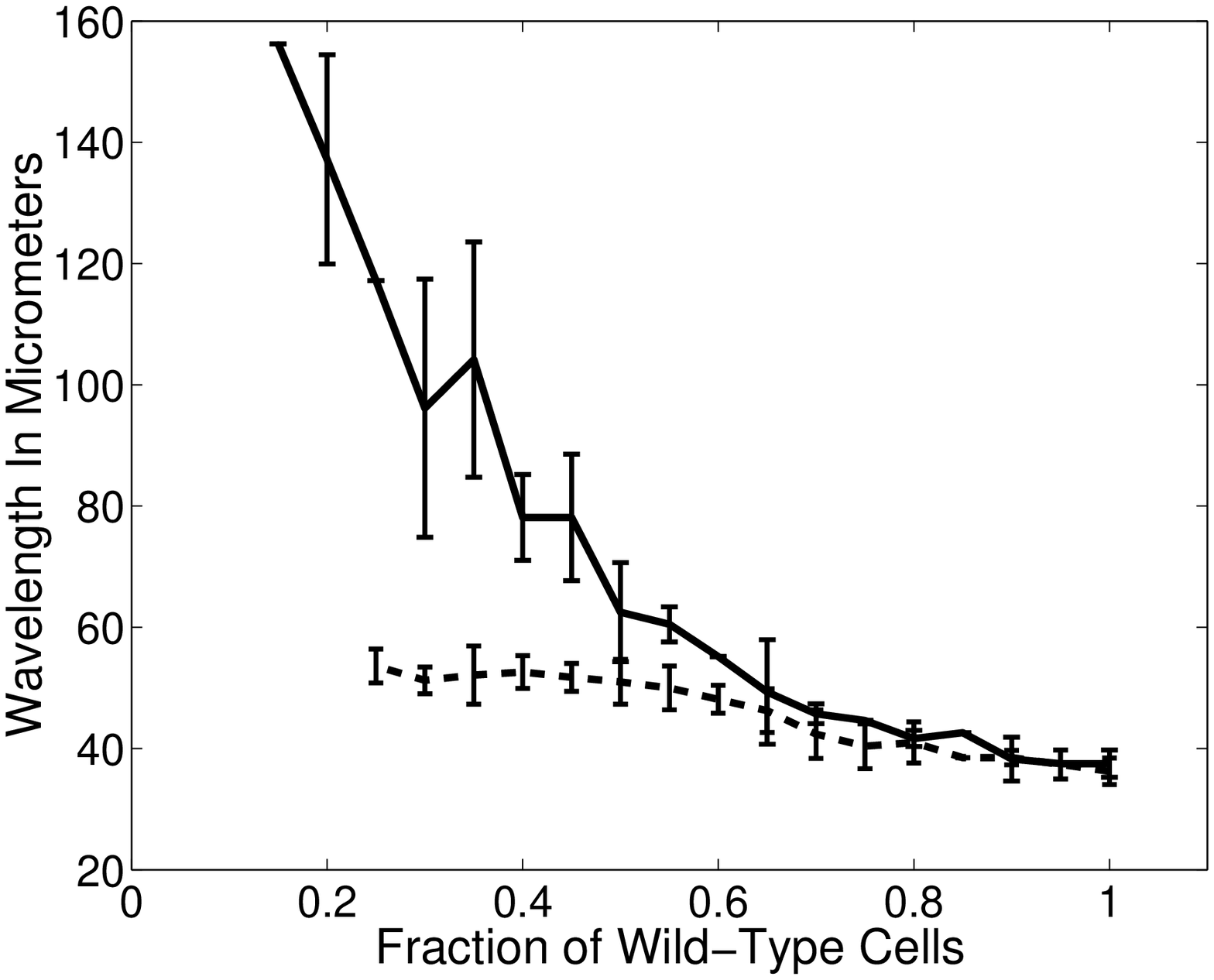}{ (a) }
\includegraphics[height=1.8in]{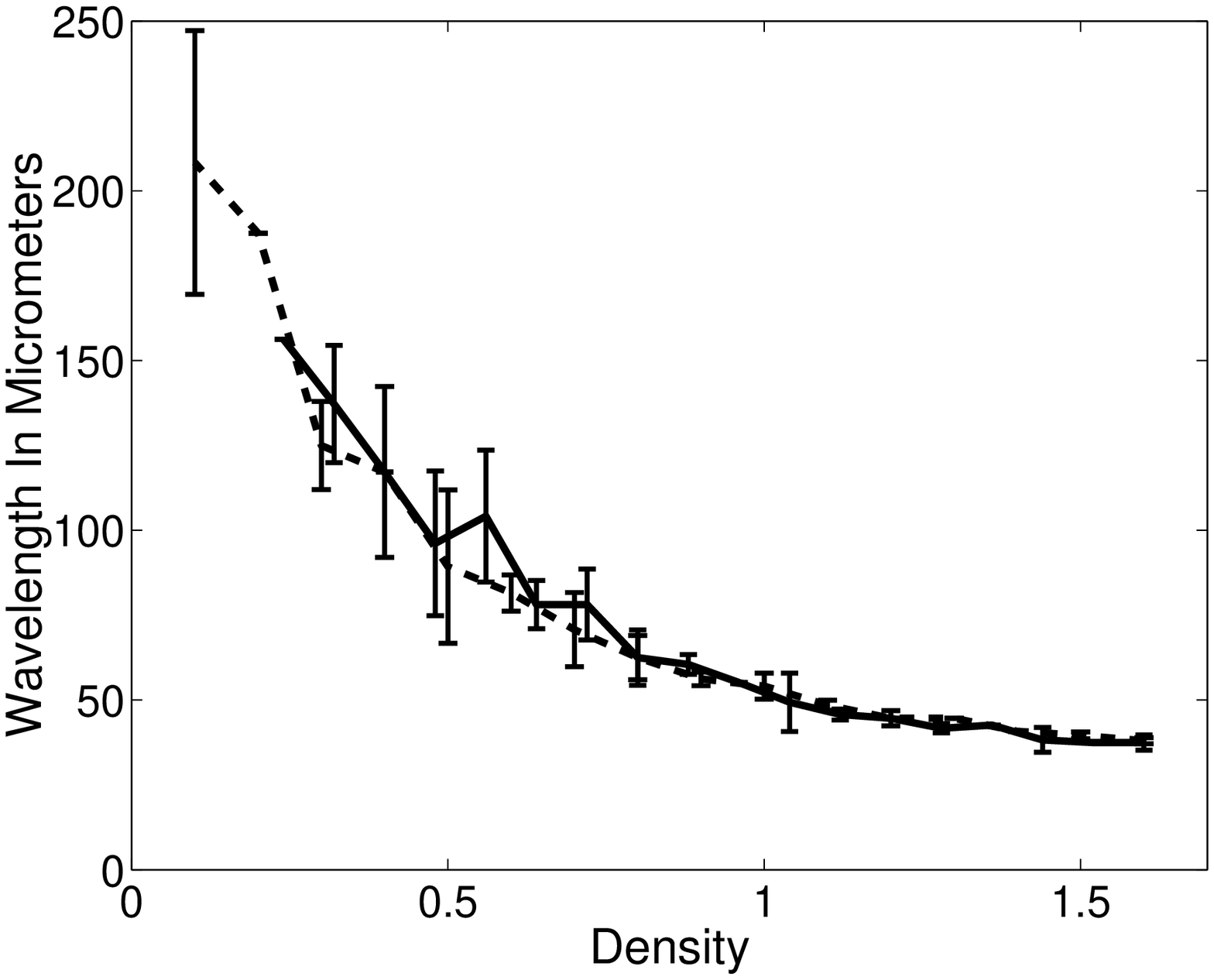}{ (b)}
\end{center}
\caption{\label{Dwavelength.eps} a) Wavelength in micrometers versus
  the fraction of wild-type cells with (dotted line) and without
  (solid line) a maximum oscillation period. b) Wavelength in
  micrometers versus wild-type density with no csgA-minus cells
  (dotted line) and when the the density of csgA-minus cells is
  increased so that the total cell density remains $1.6$ (solid
  line). Density is total cell area over total lattice area and there
  is no maximum oscillation period. For a) and b), cell length $=4$,
  $R=8$, $t=12$, $\tau=20$ (maximum oscillation period) or $\tau=2000$
  (no maximum oscillation period).}
\end{figure}

Sager and Kaiser \cite {sager1994} diluted C-signaling (wild-type)
cells with non-signaling (csgA minus) cells that were able to respond
to C-factor but not produce it themselves. When a collision occurs
between a signaling and a non-signaling cell, the non-signaling cell
perceives C-factor (and the collision), whereas the C-signaling cell
does not receive C-factor and behaves as though it has not collided.
The ripple wavelength increases with increasing dilution by
non-C-signaling cells. Simulations of this experiment with and without
an internal timer with a maximum oscillation period give very
different results. Figure \ref{Dwavelength.eps}a shows that the
dependence of wavelength on the fraction of wild type cells resembles
the experimental curve only when there is no maximum
oscillation period assumed in the model (compare with Figure 7.G
in \cite {sager1994}.) Thus our model predicts that rippling cells do
not ripple with a maximum oscillation period.  Notice that the
range of wavelengths when there is no assumed maximum oscillation
period is in good quantitative agreement with that of experiment
(compare Figure \ref{Dwavelength.eps}a, solid line with Figure 7.G
in \cite {sager1994}.

Igoshin {\it et al.} \cite{igoshin} have previously reproduced
the experimental relationship between wavelength and dilution with
non-signaling cells (see \cite{igoshin} Supplemental materials, Figure
8) by adjusting their original internal timer.  As the density of
C-signal decreases, the phase velocity slows linearly and the maximum
oscillation period of the internal timer increases continuously.
Thus, the maximum oscillation period varies in their model.  We assume
 a constant maximum oscillation period, which is either present or
 absent (longer than the simulation running time).  Note that if
the maximum oscillation period increases sufficiently with decreased
density of C-factor so that a cell is always stimulated to turn before
the internal timer would regulate a turn, then the addition of an
internal timer is superfluous.  In this case, the two models are
similar.

Our simulations show ripple wavelength increases with increased
dilution by non-signaling cells. Since wavelength also
increases with decreasing density of signaling cells (Figure
\ref{RDwavelength.eps}b), we ask if the mutant cells have any
effect on the rippling pattern.  Figure \ref{Dwavelength.eps}b shows
the wavelength dependence on the density of signaling cells when only
signaling cells are present (dotted line) and for a mixed population
of signaling cells of the same density with non-signaling cells added
so that the total cell density is always $1.6$ (solid
line).  Apparently, the decrease in C-factor explains the increase in 
wavelength. The non-signaling mutants do not affect the pattern at
all. 

\begin{figure}[ht]
\begin{center}
\includegraphics[height=2.6in]{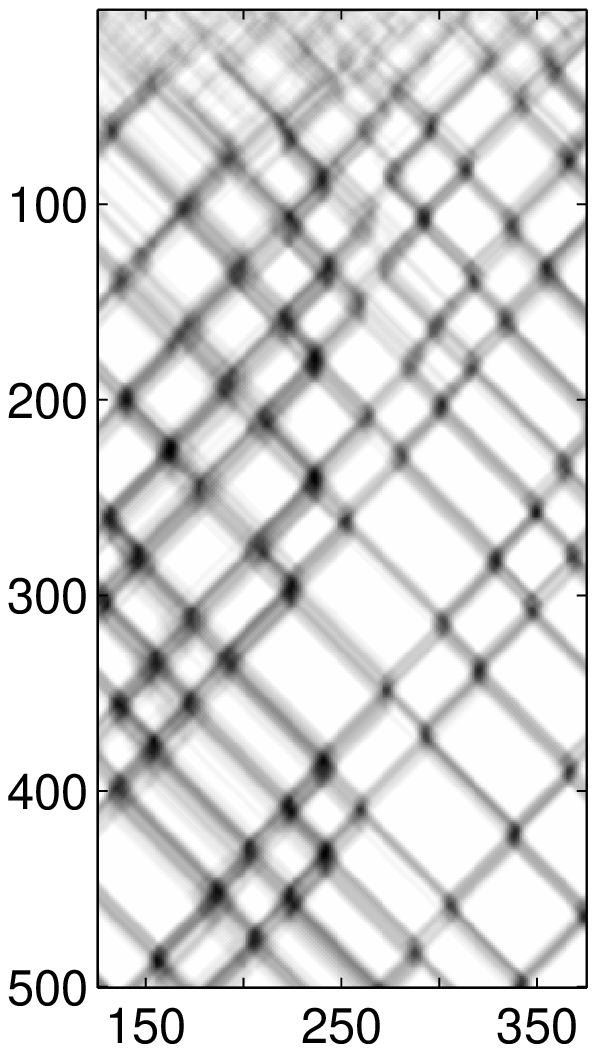}{ (a) }
\includegraphics[height=2.6in]{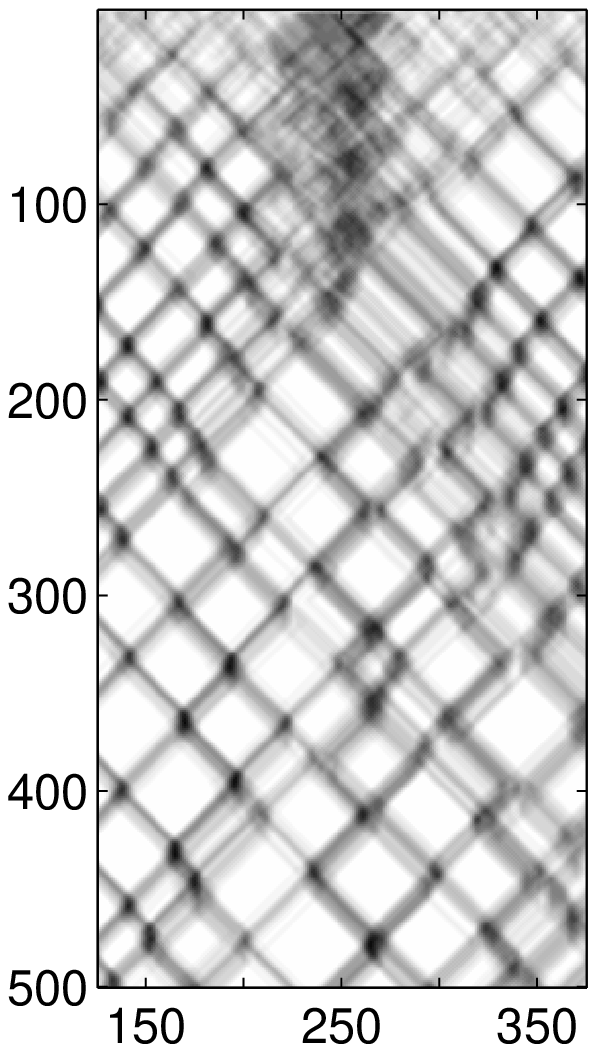}{ (b) }
\includegraphics[height=2.6in]{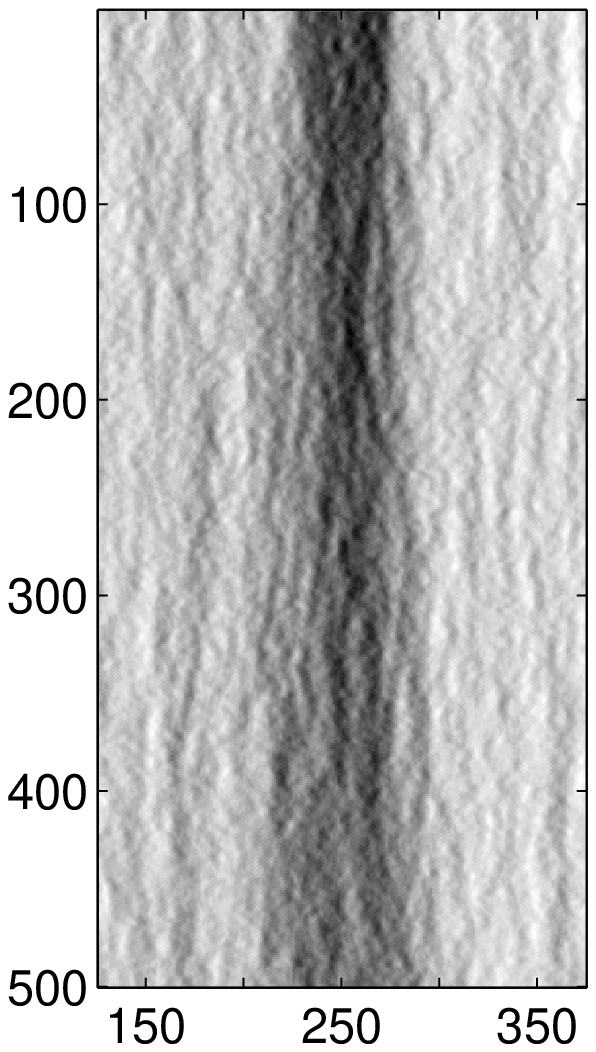}{ (c)}
\end{center}
\caption{\label{solitons2}
Cell density over  a subsection of the third row of a $5\times500$ lattice over the
first 500 time-steps for different initial conditions. Time increases
as the vertical axis descends. a) Cells are initially randomly
distributed with density 3. Cell length = 5 nodes, $R=10$ nodes,
$t=15$ nodes and $\tau=2000$. b) Same as in a), but with a central
stripe of density 15 and width 50 initially added vertically down the
lattice. c) Same as in b), but cells are assigned random orientations
at every time-step.}
\end{figure}

As density increases, wavelength decreases and the larger number of
cells are distributed over a greater number of ripples. This result is
further evidence of the the role C-signaling plays as a
density-sensing and density-modulating mechanism. To test this
further, we ran a simulation for initial conditions in which a high
density stripe stretches vertically down a lattice. As ripples formed
and propagated, the cells were quickly distributed more evenly over
the lattice (Figure \ref{solitons2}b). The redistribution of
cells occurs much faster than if cells moved randomly at each
time-step (compare Figure \ref{solitons2}, b and c). Thus, although
there is no net transport of cells larger than one wavelength when
cells are evenly distributed \cite{sager1994}, there is net migration
of cells away from high density regions.

\subsection{Discussion}

In our simulations, high-density waves of cells form from a homogeneous 
distribution of cells for a wide range of parameters and initial cell
density.  At low cell density when there is no assumed maximum
oscillation period, the wavelength between ripples is large.  
The explanation for this is that a larger region of the lattice must
be ``swept'' to collect an aggregate of cells with sufficiently high
density to reverse another aggregate. In the extreme case where
density is nearly zero, a single cell will keep traveling without ever
encountering sufficient collisions, and the wavelength is
infinite. Thus, the mechanism of rippling may be viewed as a
mechanistic ``sweeping" of an arbitrarily large area, in which cells
modulate the area that they span between reversals so as to efficiently collect
ripples of a minimum density.

At very high density, by the same argument, wavelength is small.
The number of ripples per area increases so that the number of cells per
ripple does not increase linearly with the increase in density, but less.
Nevertheless, ripples formed from high density initial conditions do
result in ripples which are wider and more dense than
ripples formed from low density initial conditions. This may not be viewed
as a limitation in design, but as the foundation of another possible
role for rippling in myxobacteria: the even distribution of cells over
an arbitrarily large area.  When there are both high and low cell
density regions (as in the simulation of Figure \ref{solitons2}b),
many high-density waves will form in the region of highest density,
and fewer, lower density waves will form in the regions of lower
density. At the interface between these regions, a high density wave
encounters a lower density wave. The high density wave will reverse
all the cells of the low density wave such that all the cells from the
low density wave return toward the low density region. The cells of
the low density wave, on the other hand, will not be able to reverse
all the cells of the higher density wave. Rather, the lower density
wave will only be able to reverse a proportional number of cells in
the high density wave. The surplus cells in the high density wave will
continue without reversing into the low density 
region. Thus, the rippling mechanism in a region of variable cell
density creates a ``pulse'' of surplus cells within the high density
region which is efficiently directed into lower density regions. 

In summary, rippling is an efficient mechanism for both forming evenly
spaced accumulations of high cell density, and evenly spaced
accumulations of nearly equal cell density. In experiments, cells do
not reflect by exactly 180 degrees. However, since most cells move
roughly parallel to each other, models based on reflection are
reasonable approximations. Modeling the experimental range of cell
orientations would require a more sophisticated CA since LGCA require
a regular lattice which does not permit many angles. In the
aggregation section below we describe a model on a triangular lattice 
which could be adjusted to incorporate 120 degree reversals.  
Although a cell is ready to turn when the
internal timer $\phi$ is greater than $\tau$, it may not be able to
turn if the opposite channel is already occupied. This is another
limitation of our LGCA model. We handle this situation by continuing
to transport the cell in its direction of orientation at each
time-step until the opposite channel is available. The effect of this
delay is negligible, even at high densities within a ripple, when
cells are so long that the probability of two cell centers occupying
the same node is very small. 

\section{A preliminary model for aggregation}

Rippling is an intermediate, transient stage of fruiting body
formation, which is not necessary for aggregation formation
\cite{kaiser_private}. Figure \ref{ripples} shows a field of
aggregation centers surrounded by ripples. In this section we  
present a different LGCA model based on C-signaling alignment. This
LGCA model reproduces the sequence and geometry of the non-rippling
stages of fruiting body formation in detail, demonstrating how
C-signaling-based alignment can account for these patterns with very
few additional assumptions.

The non-rippling stages of fruiting body formation include alignment,
streaming and aggregation. During alignment, cells form aligned
patches from a random distribution. While streaming, cells form long
chains which move cooperatively into aggregation center
\cite{jelsbak2000}. Aggregation is the phase in which cells form
rounded collections that may either recede or mature into fruiting
bodies. We model aggregation including only a simple local rule for 
C-signal-based alignment. The aggregates formed in our model are
not species-specific and do not include local rules for rippling.

\begin{figure}[ht]
  \begin{center}
    \includegraphics[height=1.5in]{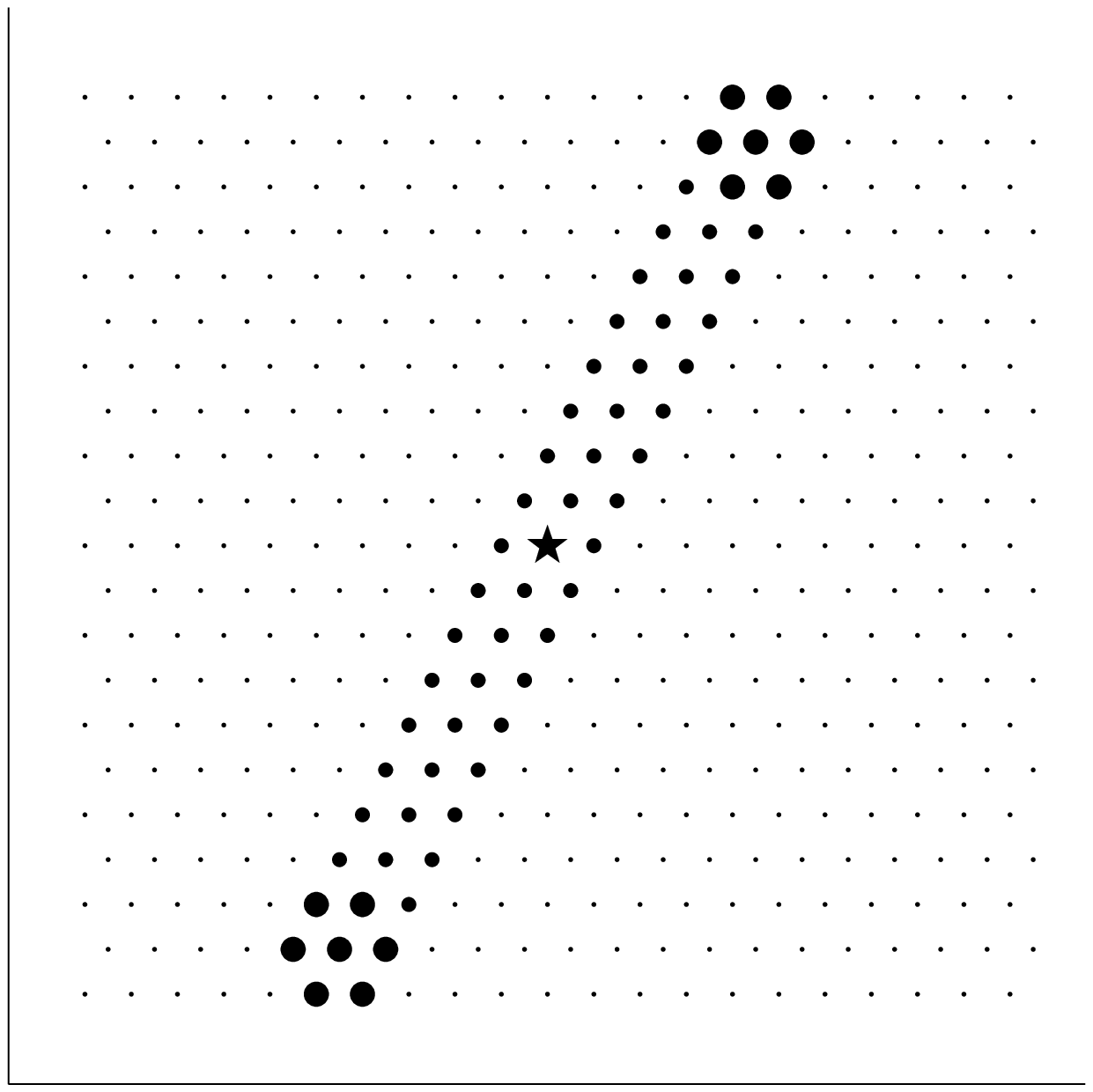}{ (a) }
    \includegraphics[height=1.5in]{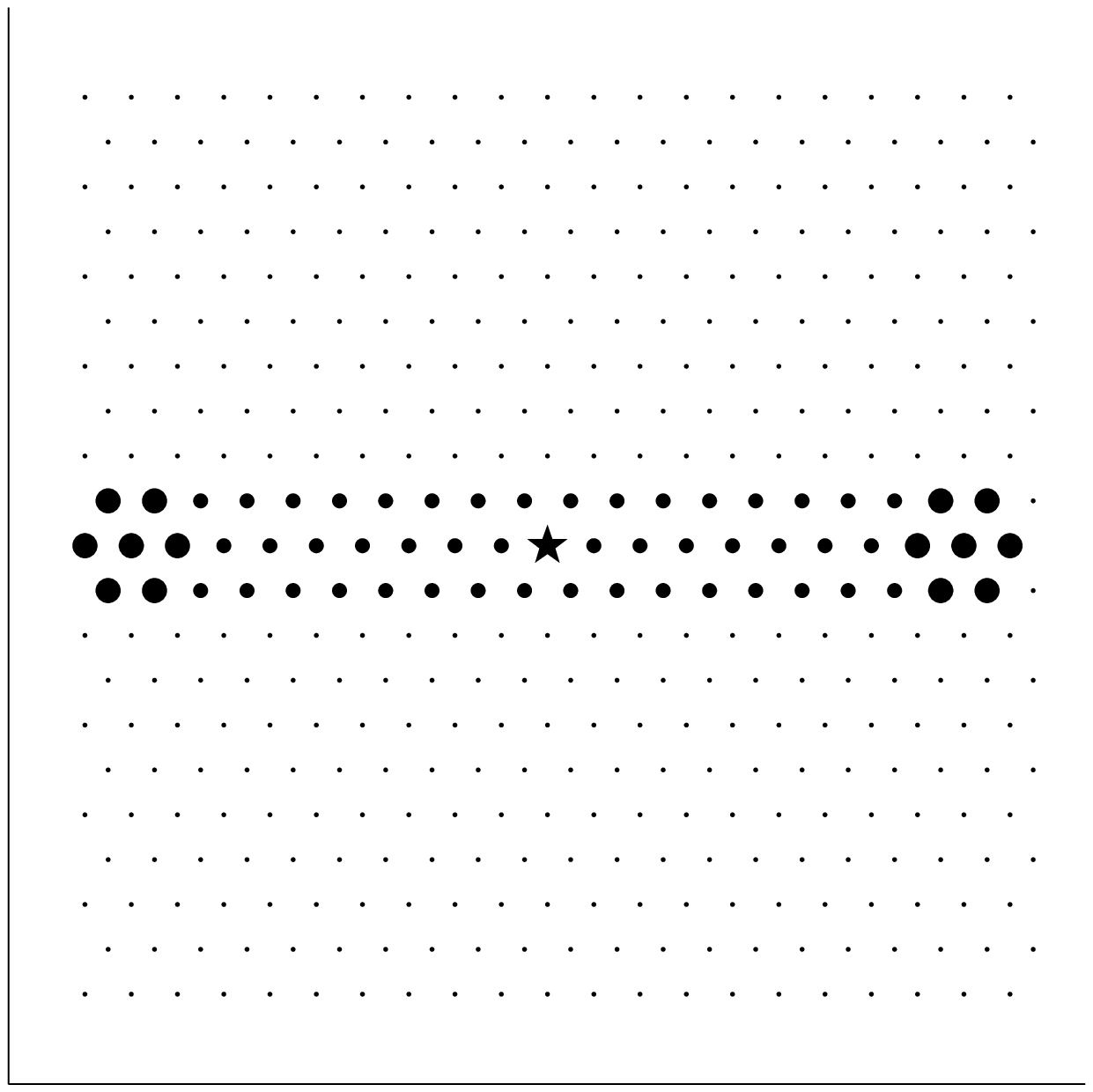}{ (b) }
    \includegraphics[height=1.5in]{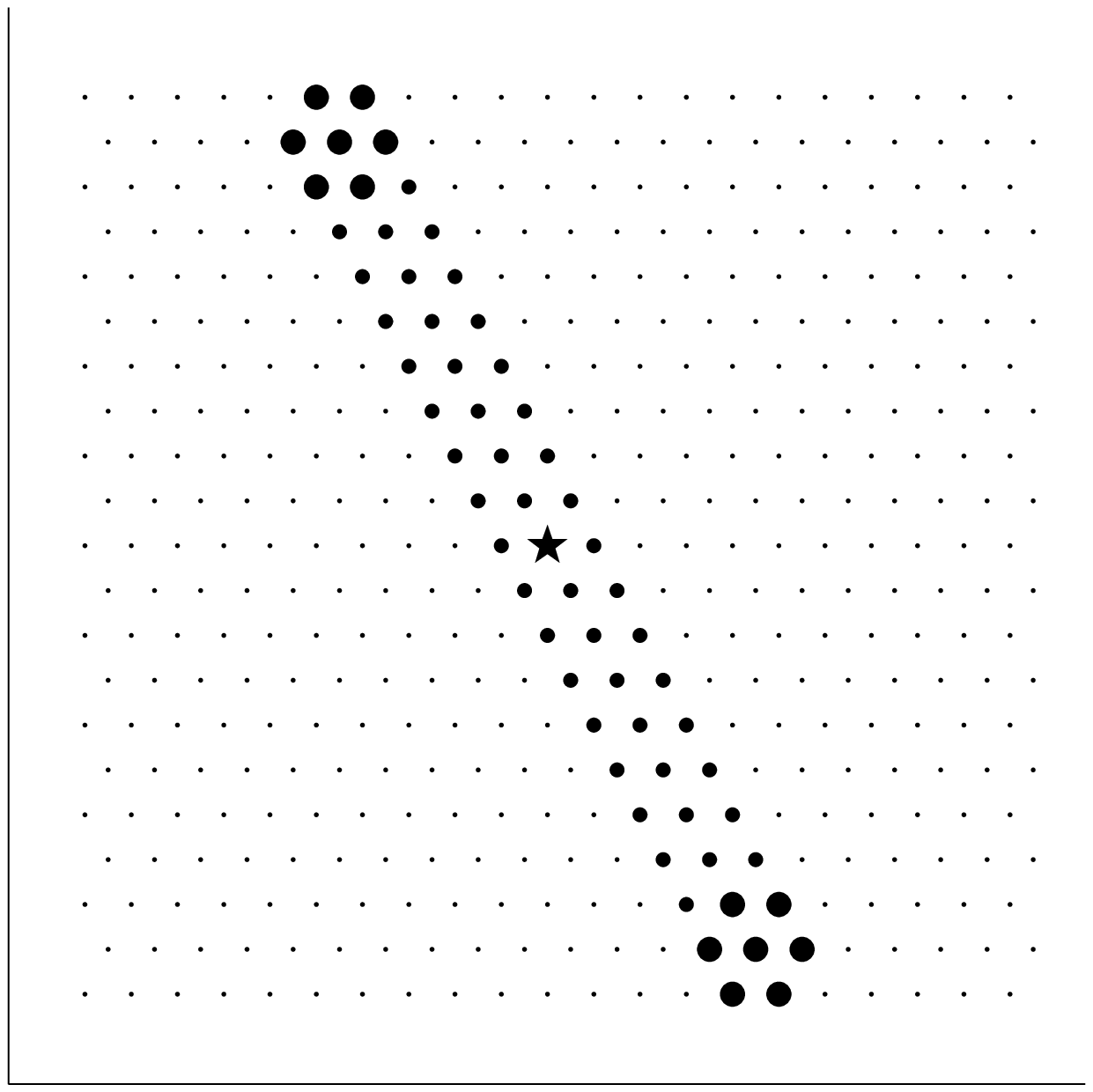}{ (c) }
  \end{center}
  \caption{\label{nbhds2} { The cell shapes of a) a cell
   oriented 60 or 240 degrees, b) a cell oriented 0 or 180 degrees
    and c) a cell oriented 120 or 300 degrees. All cells are $3 \times
    21$ for a $1 \times 7$ aspect ratio. Each cell's ``center of mass"
    is indicated by a star and the nodes of the interaction
    neighborhood where C-factor is exchanged are indicated by the
    larger black disks at the cell poles.}}
\end{figure}

We use a hexagonal lattice since cell motion during aggregation
is not one-dimensional as in rippling. In this specialized LGCA model,
identical rod-shaped cells are all modeled as $3 \times \ell$
rectangles with C-signal interaction neighborhoods as depicted in
Figure \ref{nbhds2}.  Cells move exactly one node per time-step in the
direction of their orientation and there may only be one cell center
per channel per node. In contrast to rippling, we find that the cell
aspect ratio is an important parameter for streaming and aggregation,
the simulations presented here all have a $1 \times 7$ aspect ratio
for cells. 

Myxobacteria align when they move. We choose an alignment based on
C-signaling. We use a Monte Carlo process, in which cells turn 60
degrees clockwise, 60 degrees counter-clockwise or persist in their
original direction with probability favoring directions that maximize
overlap of C-signaling nodes. Interaction only occurs when the
C-signaling nodes at the head of a cell overlap with the C-signaling
nodes at the tail of another cell, and interaction occurs regardless
of cell orientation. Head and tail C-signaling neighborhoods are shown
in Figure \ref{nbhds2}. 

\begin{figure}[ht]
  \begin{center}
    
   \includegraphics[height=1.5in,width=1.5in]{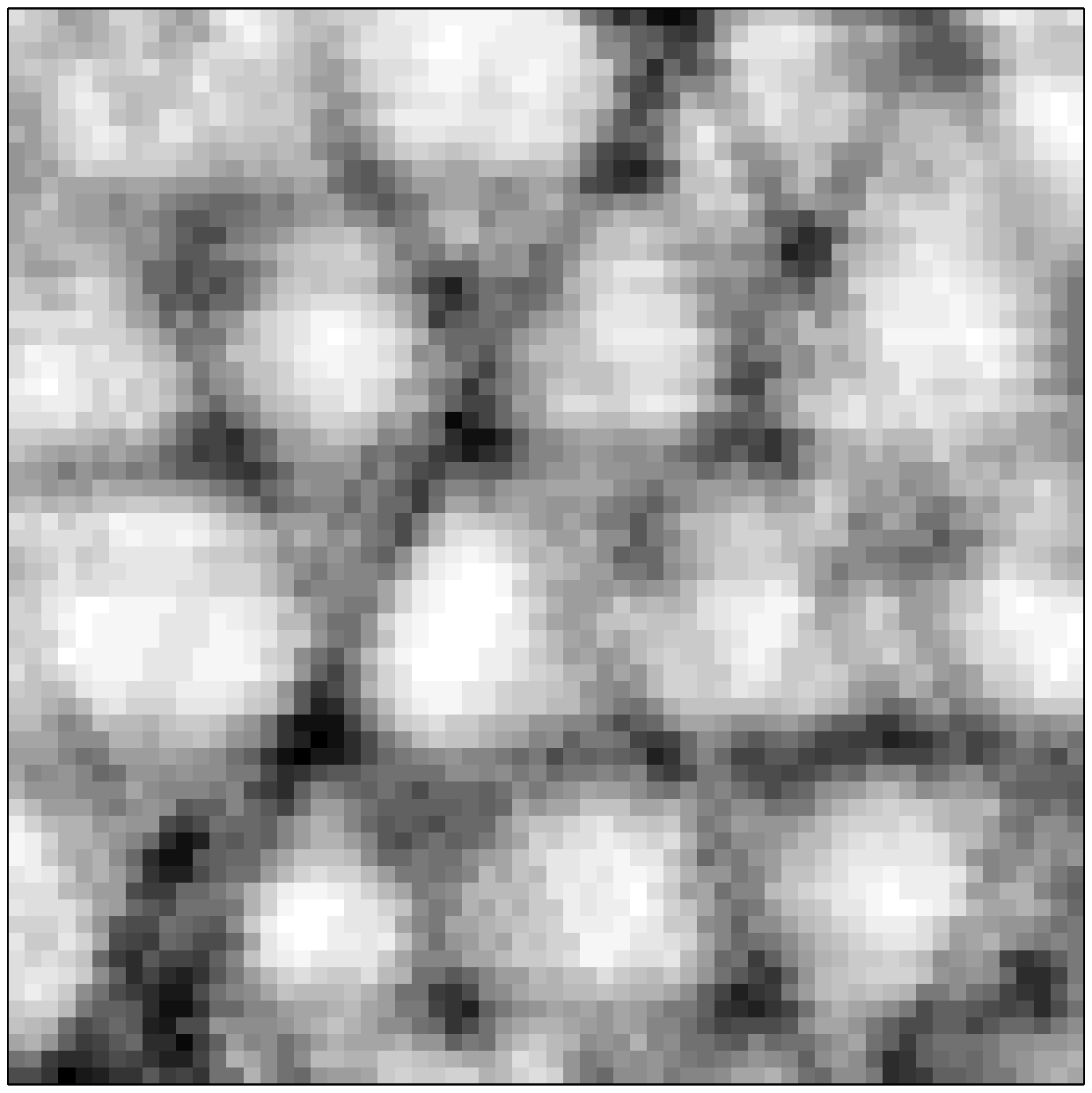}{(a)}  
   \includegraphics[height=1.5in, width=1.5in]{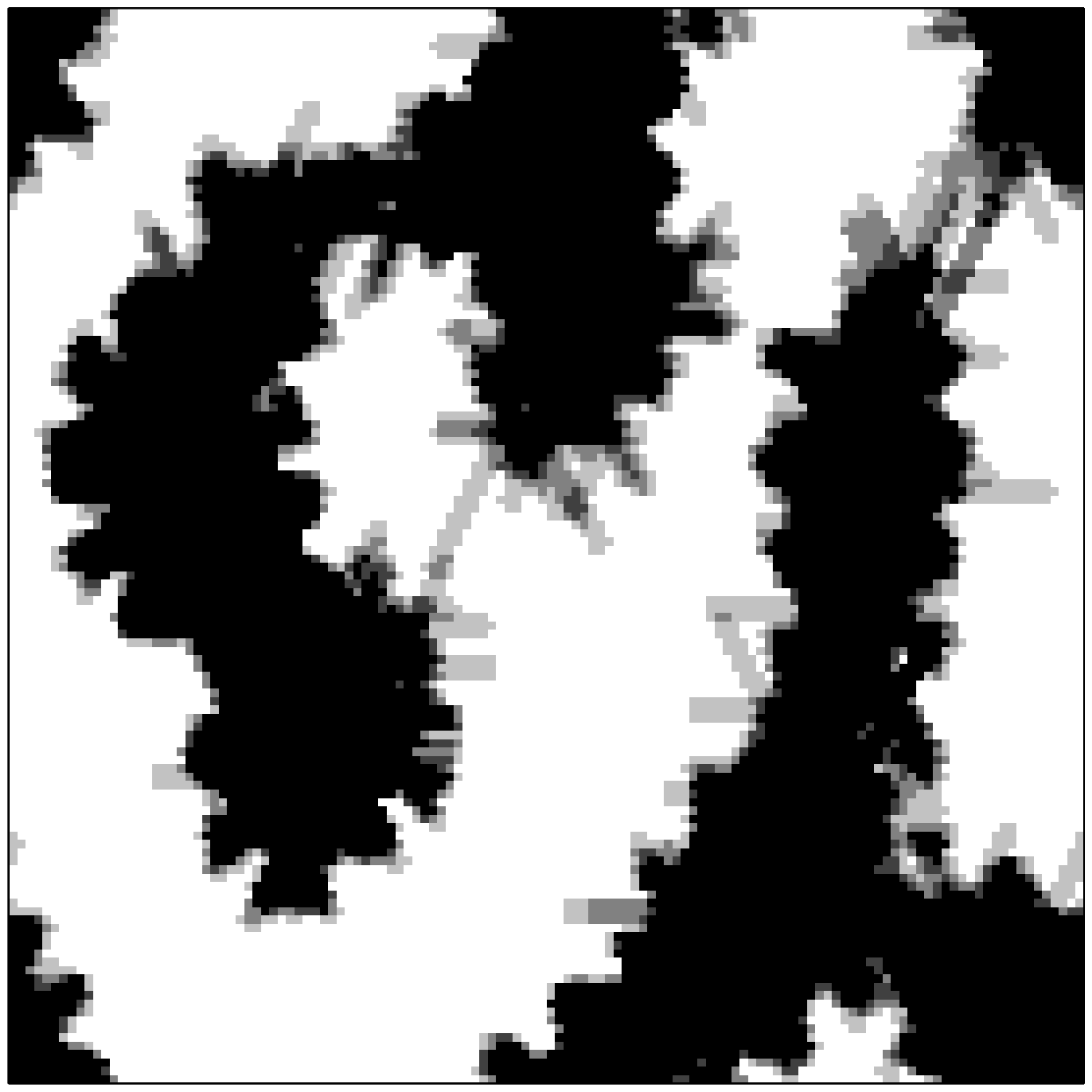}{(b)}
   \includegraphics[height=1.5in, width=1.5in]{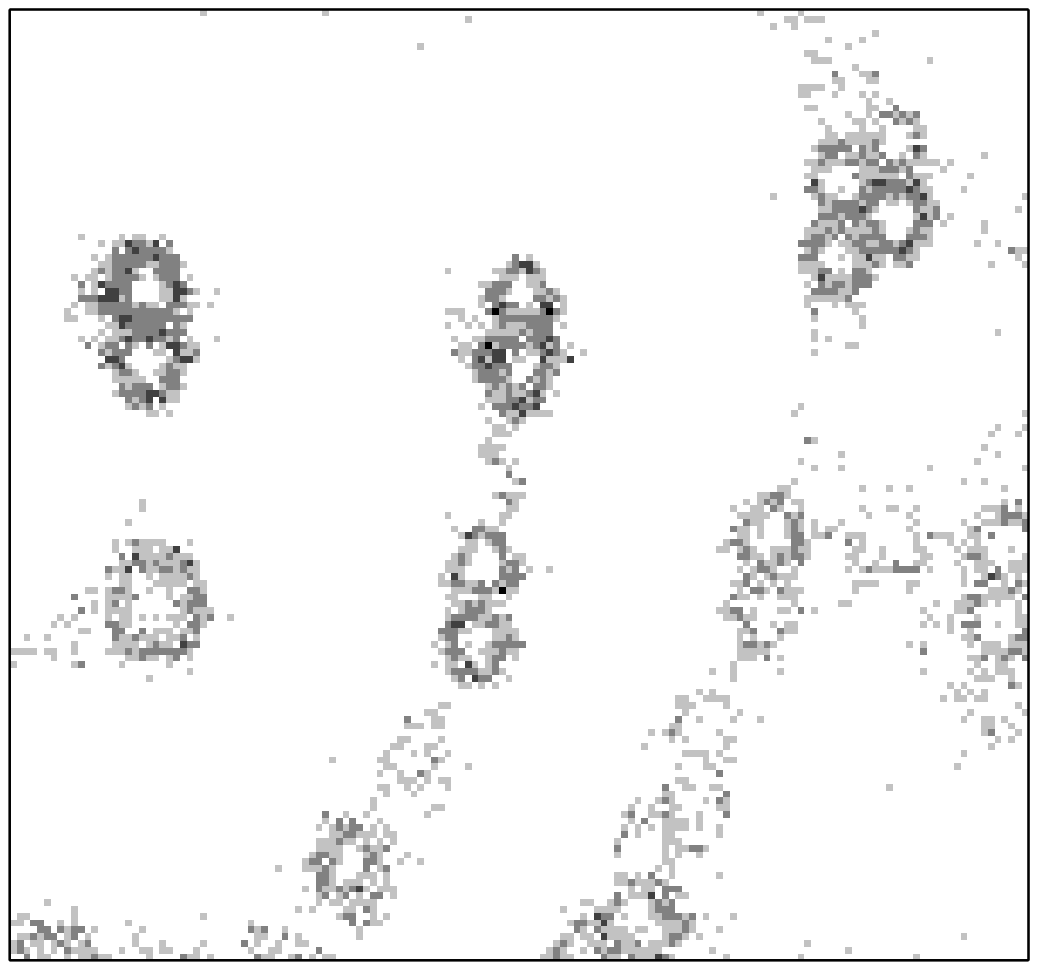}{(c)}
   \newline
   \includegraphics[height=1.5in, width=1.5in]{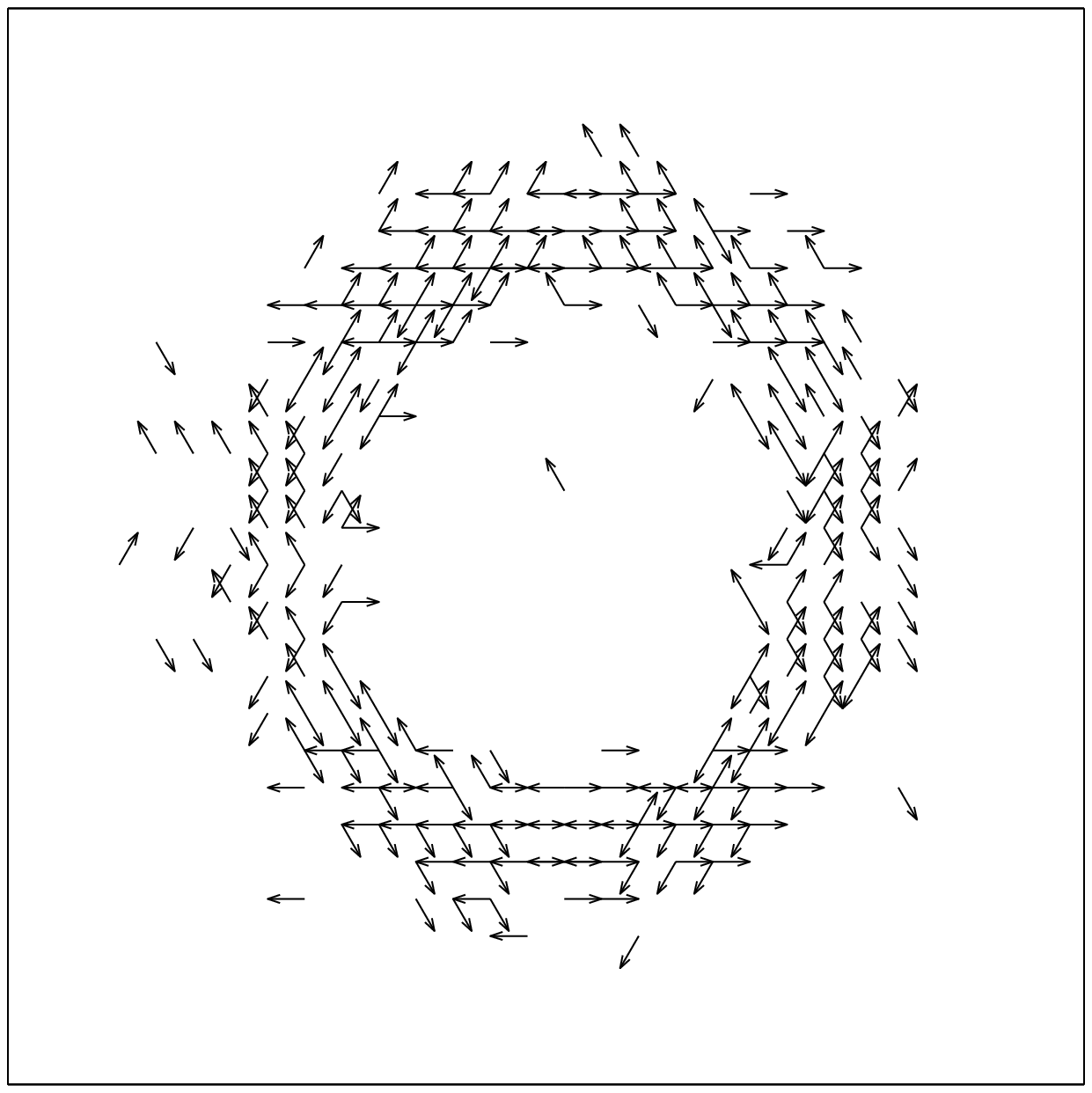}{(d)}
   \includegraphics[height=1.5in, width=1.5in]{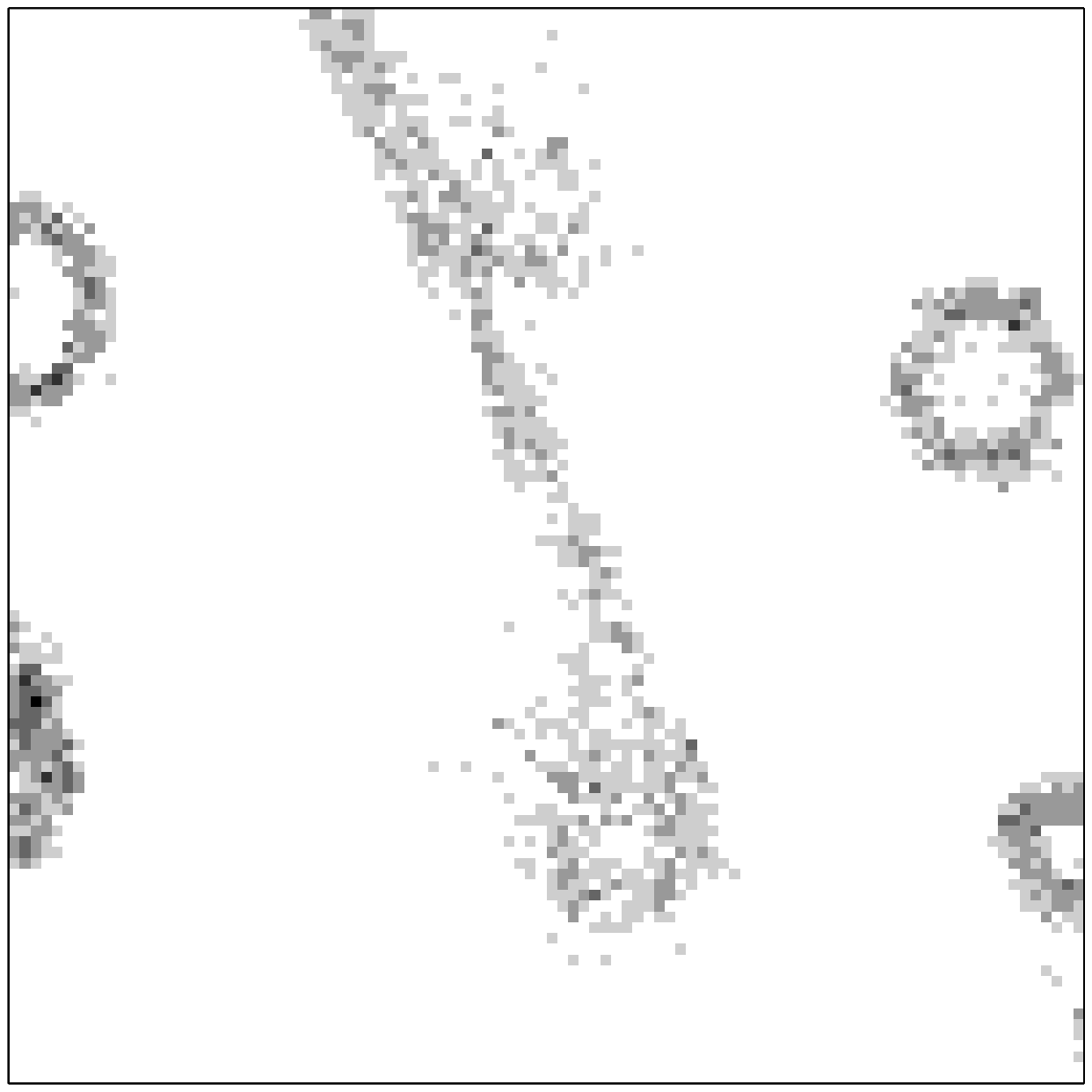}{(e)}
   \includegraphics[height=1.5in, width=1.5in]{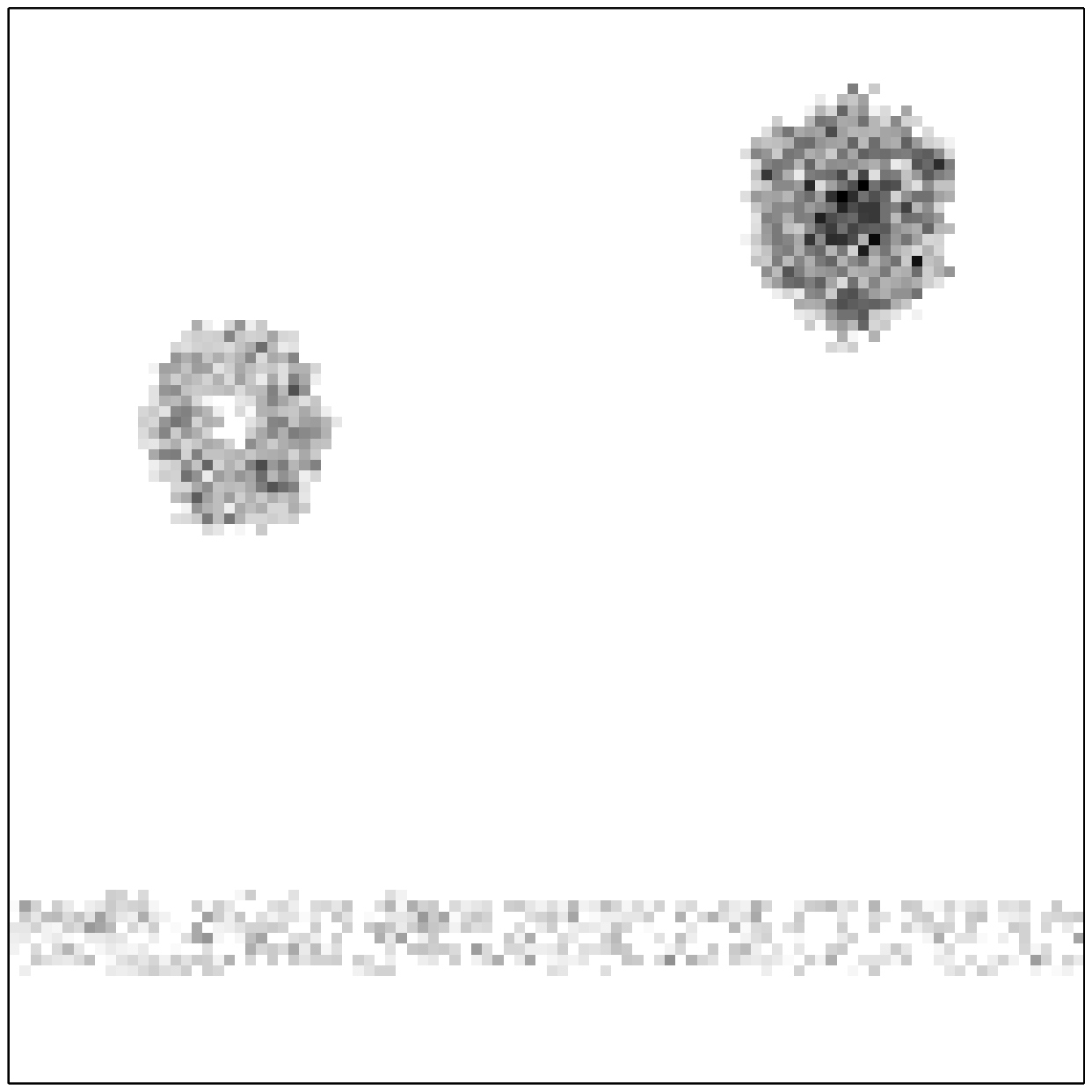}{(f)}

  \end{center}

  \caption{\label{stages} {Cell density development by C-signal
   alignment on a $256\times 256$ lattice. Initial cell density is 10.
    Cell density a) at 25 time-steps ($64\times 64$ 
    lattice sub-section) and b) at 100 time-steps ($128\times 128$
    lattice sub-section). Cell centers c) at 200 time-steps
    ($150\times 150$ lattice sub-section),  d) of a $24\times 24$
    lattice sub-section with arrows indicating direction (450
    time-steps), e) at 450 time-steps ($100\times 100$ lattice
    subsection) and  f) at 2000 time-steps ($100\times 100$ lattice
    sub-section).}}

\end{figure}

We model C-signal alignment on a $256\times 256$ lattice, in which our
initial conditions are a random distribution of cells at high
density. Within a few time-steps, cells form aligned patches (see
Figure \ref{stages}a). This reproduces the initial alignment stage
of myxobacteria cells during fruiting body formation in which cells
form an aligned patchwork \cite{kaiser1999}. Since there are only 6
directions permitted on the lattice,  the aligned patchwork appears as
a very regular triangular network. Cells are aligned both parallel and
anti-parallel within each patch since the overlap of C-signaling nodes
is maximized when cells are aligned regardless of their
orientation. This geometry is significant because cells have naturally
formed the aligned bi-directional arrays necessary for rippling. The
local rules for rippling have been suppressed in this preliminary
model, however, to evaluate the patterns formed by C-signaling-based
alignment alone.

The homogeneous triangular network is not stable over time. As cells
move, they turn and flow from one patch to
another. Cells moving in a low density area are likely to turn into a
higher density patch, so the network of cells condense into thick
aggregates (Figure \ref{stages}b). In experiments, a
myxobacteria stream will often merge into an adjacent myxobacteria
  stream \cite{stevens,mcbride}. Figure \ref{stages}b shows
that the cells move into the aggregates along streams directed into
the aggregate, reproducing the streaming stage in which  myxobacteria
cells form long aggregates that move cooperatively \cite{kaiser_private}.

The aggregates continue to condense while arranging and dividing
into many small, circular orbits about 1.5 cell lengths (about 10-20
$\mu$m) in diameter (Figure \ref {stages}c). The aggregates often
form in clusters of two or three closed orbits (Figure \ref{stages}b), 
corresponding to fused aggregates (sporangioles).  In {\it
  Stigmatella erecta}, several fruiting bodies may form in groups and 
fuse \cite{reichenbach1993}. A magnified picture of the cell centers
of a typical aggregate show that cells are arranged in dense,
concentric layers tangent to a relatively low-density inner region
(Figure \ref{stages}d).  Thus, they are geometrically equivalent
  to the basal region of aggregates in {\it Myxococcus xanthus}.

Cells in our simulation simultaneously move both clock-wise and
counter-clockwise around the aggregate, as they do in the fruiting
bodies of {\it Myxococcus xanthus} \cite{sager1993}.  The microscopic
circular or elliptic orbits of {\it Stigmatella} spp. often disappear
as cells spiral away from the aggregate \cite{white1993}.  Similarly
most orbits in our simulation  also eventually disappear as cells
spiral away from the aggregate. Nevertheless, orbits typically survive
for several hundred time-steps, which is about 5 to 10 complete
revolutions for each cell.

During myxobacteria aggregate formation, several aggregation centers
will form and, inexplicably, one aggregation center will grow as an
adjacent aggregation center disappears \cite{kaiser_private}.  Our
simulations offer a closer look at this process: a stream may form
that connects two adjacent aggregation centers and, stochastically,
cells stream from one aggregation center to another until the largest
aggregation center absorbs the smaller one (Figure \ref{stages}e).

Figure \ref{stages}f shows several stable aggregates which have
developed at 200 time-steps. Notice that the stable hollow aggregate
has a much thicker annulus of cells than the non-stable orbits of
Figure \ref{stages}c, suggesting that only large hollow aggregates
are stable. In our simulation at a threshold density within the
aggregate, the hollow region of an aggregate will fill with cells such
that cells are are arranged in six dense overlapping layers (Figure
\ref{stages}f).  The third aggregate shown is a chain of cells which
span the lattice and thus form an orbit due to the periodic boundary
conditions of the lattice.

It has been proposed in \cite{white1993} that circular motility at
aggregation sites, trail following and local accumulation of slime
account for fruiting body formation in {\it Stigmatella} spp. We
hypothesize that once C-signaling has drawn cells into an aggregate,
cell and slime cohesivity cause myxobacteria cells to round up into a
mound while constant cell velocity pushes cells toward the surface of
the mound, so that cells form a dense hemisphere of cells spiraling
around a hollow center. In our model, a closed, circulating orbit of
cells is the only stable configuration of a stationary aggregate since
cells are constantly moving. At low and intermediate density, these
orbits are hollow and cells are arranged tangentially within an
annulus. At a threshold density, however, every channel of every node
within the aggregate becomes occupied and there is no hollow
center. We hypothesize that in a more advanced three-dimensional
model, the addition of a local rule accounting cell and slime
cohesivity will cause cells to round while maintaining the hollow
center. 

This model is preliminary because only C-signal based alignment is 
modeled and the aggregates formed are not species-specific.  In
this simulation, a 256x256 lattice size was chosen, which corresponds
to an 80x80 $\mu$m region and about 10,000 cells for an averaged cell
density of 10, much smaller compared to normal fruiting bodies that 
may be up to 1000 $\mu$m in diameter. Aggregation in this model is
described in more detail in \cite{submitted_aggregation}. 

This model can be adjusted to
model rippling and aggregation concurrently by incorporating the local rules
for rippling described in Section 3.2. Tracking of rippling
cells in experiment (see Figure 6 in \cite{sager1994} and Figure 6 in
\cite{sager1993}) suggest that reversals are about 150 degree
changes in orientation rather than exactly 180 degrees, as we have
assumed in the model for rippling of this paper. On a triangular
lattice, a reversal of 120 degrees would be an equally good
approximation of the 150 degree rotation. C-signaling based alignment
combined with local rules for rippling would ensure that the majority
of cells still remain parallel. Nevertheless, a regular shift in
orientation by 120 {\it or} 180 degrees may have an interesting effect
on the final distribution of cells and, subsequently, fruiting
bodies. 

Fruiting bodies among different myxobacteria species are very
diverse (see Figure 3 in \cite{pfister1993}).  For example, while
{\it Myxococcus} fruiting bodies are a relatively simple,  single
mound of cells, other species form clusters of mounds called
sporangioles that are raised on a stalk.  The interaction of
adventurous verses social motility may account for these different
morphologies \cite {kaiser_private}. During stalk formation of
{\it Stigmatella} spp., cells are arranged perpendicularly to the
mound as the fruiting body is lifted \cite
{white1993,vasquez1985}. Also, fruiting body stalks may be
composed of a larger, second cell-type \cite{white1993}. Thus,
once the stages of fruiting body formation have been modeled in
general, it would be interesting to determine which parameters
need to be varied to model the fruiting bodies of different
species.

\section{Summary}

In this paper, we present a new LGCA approach for modeling cells
which is computationally efficient yet approximates continuum
dynamics more closely than assuming point-like cells.  As an
example of this new approach, we present a model for myxobacteria
rippling based on the hypothesis of precise reflection. The results of
our model show that rippling is stable for a wide range of parameters,
C-signaling plays an important role in modulating cell density during
rippling, and non-C-signaling cells have no effect on the rippling
pattern when mixed with wild-type cells. Further, by comparing
model results with that of experiment, we can conclude reversals
during rippling would not be regulated by a built-in maximum
oscillation period. We also present a second LGCA model based on 
C-signal alignment which reproduces the sequence and geometry of
the non-rippling stages of fruiting body formation in detail,
showing that a simple local rule based on C-signaling can account
for many experimental observations.

{\bf Acknowledgments}

We would like to thank Dale Kaiser, Frithjof Lutscher and Stan
Mar\'{e}e for very helpful discussions. MSA is partially supported by
grant NSF IBN-0083653. YJ is supported by DOE under contract
W-7405-ENG-36. MAK is supported by the Center for Applied Mathematics
and the Interdisciplinary Center for the Study of Biocomplexity,
University of Notre Dame, and by DOE under contract W-7405-ENG-36.

\end{document}